\documentclass[fleqn,usenatbib]{mnras}

\usepackage{newtxtext,newtxmath}
\usepackage[T1]{fontenc}

\DeclareRobustCommand{\VAN}[3]{#2}
\let\VANthebibliography\thebibliography
\def\thebibliography{\DeclareRobustCommand{\VAN}[3]{##3}\VANthebibliography}

\usepackage{color}
\usepackage{soul}
\usepackage{bm}
\usepackage{orcidlink}

\usepackage{graphicx}	% Including figure files
\newcommand{\cma}{Comput. Math. Appl.}

\newcommand{\cpc}{Comput. Phys. Commun.}
\newcommand{\jgrsp}{J. Geophys. Res. Space Phys.}
\newcommand{\jpp}{J. Plasma Phys.}
\newcommand{\lrsp}{Living Rev. Sol. Phys.}
\newcommand{\pof}{Phys. Fluids}
\newcommand{\pop}{Phys. Plasmas}
\newcommand{\scts}{Sci. China E Technol. Sci.}

\newcommand{\legolas}{\textsc{legolas}}
\newcommand{\amrvac}{\textsc{mpi-amrvac}}

\newcommand{\bfb}{\bm{B}}
\newcommand{\bfj}{\bm{j}}
\newcommand{\bfv}{\bm{v}}
\newcommand{\bfkappa}{\bm{\kappa}}
\newcommand{\bfk}{\bm{k}}
\newcommand{\bfx}{\bm{x}}
\newcommand{\bfum}{\bm{\mathsf{I}}}
\newcommand{\calh}{\mathcal{H}}
\newcommand{\cale}{\mathcal{E}}

\newcommand{\ey}{\hat{\bm{e}}_y}
\newcommand{\ez}{\hat{\bm{e}}_z}
\newcommand{\im}{\mathrm{i}}

\newcommand{\imag}{\mathrm{Im}}

\newcommand{\dV}{\,\mathrm{d}x\,\mathrm{d}y\,\mathrm{d}z}

\allowdisplaybreaks

\title[Tearing-thermal instability in current sheets]{The coupled tearing-thermal instability in coronal current sheets from the linear to the non-linear stage}

\author[J. De Jonghe and S. Sen]{
Jordi De Jonghe$^{1}$\thanks{E-mail: jordi.dejonghe@kuleuven.be} \orcidlink{0000-0003-2443-3903}
and Samrat Sen$^{2}$ \orcidlink{0000-0003-1546-381X}
\\
$^{1}$School of Mathematics and Statistics, University of St Andrews, Mathematical Institute - North Haugh, St Andrews KY16 9SS, United Kingdom\\
$^{2}$Instituto de Astrofísica de Canarias, Vía Láctea, E-38205 La Laguna, Tenerife, Spain
}

\date{Accepted 2024 December 08. Received 2024 December 03; in original form 2024 October 03}

\pubyear{2024}

\begin{document}
\label{firstpage}
\pagerange{\pageref{firstpage}--\pageref{lastpage}}
\maketitle

\begin{abstract}
In the solar corona, magnetically sheared structures are unstable to both tearing and thermal instabilities in a coupled fashion. However, how the choice of linear perturbation modes influences the time-scale to achieve the thermal runaway in a coupled tearing-thermal coronal current sheet is not well understood to date. Here, we model a force-free Harris current sheet under solar coronal conditions to investigate this coupling in the linear and non-linear regimes. In the linear regime, we adopt the magnetohydrodynamic spectroscopy code \legolas{} to compare the current sheet under thermal and thermoresistive conditions, after which we initialise non-linear simulations (with \amrvac{}) with the unstable, linear tearing and thermal perturbations obtained with \legolas{}. It is shown that part of the unstable thermal quasi-continuum adopts tearing properties in the linear stage, but that it is not until the non-linear stage is reached that a true thermal `runaway' effect leads to condensations inside tearing-induced flux ropes. Hence, the linear stage is governed by the dominant tearing instability whilst condensations form due to tearing-thermal coupling in the non-linear stage. Our results imply that perturbing an equilibrium current sheet with the fastest growing linear mode skips the mode mixing phase in which the dominant instability traditionally emerges, and significantly reduces the time-scale to enter into the non-linear stage and thermal runaway process from its equilibrium configuration.
\end{abstract}


%%%%%%%%%%%%%%%%%%%%%%%%%%%%%%%%%%%%%%%%%%%%%%%%%%

%%%%%%%%%%%%%%%%% BODY OF PAPER %%%%%%%%%%%%%%%%%%

\section{Introduction}\label{sec:intro}
Many events in the solar corona are strongly tied to the process of magnetic reconnection. In this fundamental process in a resistive medium, magnetic field lines break and reconnect, converting magnetic energy into thermal and kinetic energy \citep{Biskamp2000}. In the solar atmosphere, this often leads to eruptions such as coronal mass ejections \citep{Gosling1995, Antiochos1999, Karpen2012} and solar flares \citep{Giovanelli1939, Priest2000, Hesse2020}.

When a current layer in a magnetically sheared plasma is perturbed, reconnection may be triggered by the tearing instability \citep{Furth1963}, which fragments the current sheet by establishing reconnection points along the sheet. As a result, the current sheet is segmented into magnetic islands, or plasmoids. In the presence of a guide field, these magnetic islands turn into a bundle of twisted magnetic field lines wound about a common axis, and are called magnetic flux ropes \citep{Chen2011, Priest2014}. After this first instance of the tearing instability, the structure can become unstable again to secondary tearing, creating long chains of plasmoids, thus also referred to as the plasmoid instability \citep{Loureiro2007, Huang2013}.

However, the dynamics of the solar corona are not only governed by resistive effects, but depend strongly on the non-adiabatic effects of thermal conduction, radiative energy loss, and background heating. These three components form a delicate balance in the corona, where an increase in thermal energy loss cools the plasma and leads to a higher plasma density. In turn, this enhances the energy radiation again, cooling the plasma even faster and triggering a `runaway' process in which the plasma density increases rapidly and the temperature drops significantly. This is known as thermal instability \citep{Parker1953, Field1965}, and was suggested as a driving force behind the formation of prominence and arcade structures in the solar corona \citep{Smith1977, Priest1979}. Since then, the effect has been studied linearly in solar coronal conditions by \cite{VanDerLinden1991a, VanDerLinden1991c, VanDerLinden1992, Soler2011} and, more recently, non-linearly in a variety of simulations investigating prominence formation \citep{Xia2012, Keppens2014} and coronal rain \citep{Fang2013, Fang2015a, Fang2015b, Xia2017, Kohutova2020, Li2022, Sen2024}.

This overwhelming evidence for the importance of thermal instability in solar phenomena has led to a renewed interest in its fundamental properties, both linear and non-linear, and how it interacts with other magnetohydrodynamic (MHD) waves and instabilities. In \cite{Claes2019, Claes2020a}, the authors studied the interaction of slow MHD and thermal (entropy) modes, both linearly and non-linearly, and in \cite{Hermans2021}, they investigated the influence of the radiative loss function on the onset and long evolution of thermal modes. Though the interaction between tearing and thermal instability has not received much attention, the linear growth rate of the tearing instability in a pre-flare current sheet was recently shown to be modified by the non-adiabatic effects of resistivity, radiative energy loss, and thermal conduction by \cite{Ledentsov2021a, Ledentsov2021b, Ledentsov2021c}. The results were extended by \cite{Sen2022, Sen2023} to incorporate background heating, and beyond the linear evolution into the non-linear regime using 2D and 3D simulations respectively. In 2D, they found that the tearing growth rate increases by an order of magnitude when non-adiabatic terms are included, and the current sheet formed a chain of cool plasmoids, similar to prominence or coronal rain, due to the simultaneous occurrence of chaotic tearing and thermal runaway. In 3D, the tearing instability dominated the early evolution, modifying the magnetic topology forming magnetic flux ropes before the condensations are formed due to the thermal instability. The choice of the perturbations to initiate the instability in the current sheet reported in \cite{Sen2022, Sen2023} were arbitrary, where they used magnetic field perturbations in the form of multi-island structures. However, how the choice of the linear tearing and thermal modes influence the non-linear evolution of the current sheet is not well understood, and warrants a deeper investigation.

In this work, we advance the latter study to investigate the 3D evolution of the tearing-thermal evolution of a force-free coronal current sheet by performing a preliminary parameter study of the linear regime and exploiting the linear results to initiate non-linear simulations, including resistivity, (field-aligned) thermal conduction, radiative losses, and a constant and uniform background heating. The paper is organized as follows. In Sec. \ref{sec:config}, we introduce our exact configuration and methodology. The results of the linear and non-linear aspects of this study are presented in Sec. \ref{sec:results}. Finally, we formulate our conclusion in Sec. \ref{sec:discuss}.

\section{Configuration \& Methodology}\label{sec:config}
To model a solar coronal current sheet, we assume a static (velocity $\bfv = 0$), force-free Harris current sheet of width $a$,
\begin{equation}
    \bfb_0 = B_0 \left( \tanh\frac{x}{a}\ \ey + \sqrt{1 - \tanh^2 \frac{x}{a}}\ \ez \right),
\end{equation}
with $B_0 = 1$, or $2\,\mathrm{G}$ in physical units (unit magnetic field strength $\bar{B} = 2\,\mathrm{G}$), and a constant plasma density $\rho_0$, temperature $T_0$, and gas pressure $p_0$. This magnetic field strength is in line with solar coronal conditions at a height of $1.05-1.35$ solar radii \citep{Kumari2019, Yang2020}. Together with a typical length scale of $\bar{L} = 10^9\,\mathrm{cm}$ and $\bar{T} = 10^6\,\mathrm{K}$, this results in a reference density $\bar{\rho} \simeq 1.93 \times 10^{-15}\,\mathrm{g}\,\mathrm{cm}^{-3}$ for the fully ionized hydrogen plasma that we will consider here. The dimensionless equilibrium parameters are set to $a = 0.5$, $\rho_0 = 0.2$, and $T_0 = 0.5$, with $p_0 = \rho_0 T_0 = 0.1$. Hence, the plasma-$\beta$ is smaller than unity ($\beta = 0.2$), as is appropriate for the solar corona.

To study this configuration, we consider the MHD equations including resistivity, radiative cooling, background heating, and parallel thermal conduction. To quantify the natural oscillations and instabilities of this current sheet, we rely on the open-source code \legolas{} \citep[see \url{https://legolas.science}]{Claes2020b, DeJonghe2022, Claes2023}, which implements the linearised forms of the (dimensionless) equations
\begin{align}
    \frac{\partial \rho}{\partial t} = &-\nabla \cdot (\rho\bfv), \\
    \rho\frac{\partial \bfv}{\partial t} = &-\nabla p - \rho\bfv \cdot \nabla \bfv + \bfj \times \bfb, \\
    \rho\frac{\partial T}{\partial t} = &-\rho\bfv \cdot \nabla T - (\gamma - 1)p\nabla\cdot\bfv - (\gamma - 1)\rho \left( \rho \Lambda(T)-\calh \right) \nonumber\\
    &+ (\gamma - 1)\nabla\cdot(\bfkappa_\parallel \cdot \nabla T) + (\gamma - 1)\eta\bfj^2, \\
    \frac{\partial \bfb}{\partial t} = &\nabla \times (\bfv \times \bfb) - \nabla \times (\eta\bfj), \\
    p = &\rho T,
\end{align}
with $\bfj = \nabla \times \bfb$ the electric current density, $\Lambda(T)$ the radiative losses, $\mathcal{H}$ the energy gain due to heating, $\bfkappa_\parallel$ the thermal conductivity tensor along the magnetic field, and $\gamma = 5/3$ the ratio of specific heats for a monoatomic gas (fully ionized hydrogen plasma). After imposing a Fourier form for the perturbed quantities $f_1 \in \{ \rho_1, \bfv_1, T_1, \bfb_1 \}$,
\begin{equation}\label{eq:fourier}
    f_1(\bfx, t) = \hat{f}_1(x)\,\exp\left[\im \left( k_2 y + k_3 z - \omega t \right) \right],
\end{equation}
the resulting generalised eigenvalue problem is solved for the natural frequencies $\omega$ and corresponding perturbation amplitudes $\hat{f}_1(x)$ \citep[see][for a detailed overview of the methodology and code]{Claes2020b}. As we will show, the sheet is unstable to the resistive tearing instability and has an unstable thermal quasi-continuum. To investigate how these instabilities interact in the non-linear regime, each type is added to the equilibrium to act as the initial condition in a non-linear simulation with the open-source code \amrvac{} \citep[see \url{https://amrvac.org}]{Porth2014, Xia2018, Keppens2023}, where the MHD equations take the form
\begin{align}
    \frac{\partial\rho}{\partial t} = &-\nabla \cdot (\rho\bfv), \label{eq:vac-continuity}\\
    \frac{\partial (\rho\bfv)}{\partial t} = &-\nabla\cdot (\rho\bfv\bfv + p_\mathrm{tot}\bfum - \bfb\bfb), \\
    \frac{\partial\cale}{\partial t} = &-\nabla\cdot (\cale\bfv + p_\mathrm{tot}\bfv - \bfb\bfb\cdot\bfv) + \eta\bfj^2 - \bfb\cdot\nabla\times (\eta\bfj) \\&- \rho(\rho\Lambda(T) - \calh) + \nabla\cdot (\bfkappa_\parallel \cdot \nabla T), \\
    \frac{\partial\bfb}{\partial t} = &-\nabla\cdot (\bfv\bfb-\bfb\bfv) - \nabla\times (\eta\bfj), \\
    \nabla\cdot\bfb = &\ 0, \label{eq:vac-Bdivfree}
\end{align}
with $p_\mathrm{tot} = p + \bfb^2/2$ the total pressure, $\bfum$ the unit tensor, and $\cale$ the total energy density
\begin{equation}
    \cale = \frac{p}{\gamma-1} + \frac{\rho v^2}{2} + \frac{\bfb^2}{2}.
\end{equation}

In both codes, the resistivity is assumed uniform and set to $\eta = 10^{-3}$ (or $1.28\times 10^{13}\,\mathrm{cm}^2\,\mathrm{s}^{-1}$) and the parallel thermal conductivity is assumed to be of the Spitzer-type, i.e. $\bfkappa_\parallel \sim 10^{-6}\ T^{5/2}\,\mathrm{erg}\,\mathrm{cm}^{-1}\,\mathrm{s}^{-1}\,\mathrm{K}^{-1}$, along the magnetic field. For the radiative cooling, we use the \textsf{Colgan\_DM} cooling curve, which combines the results of \cite{Dalgarno1972} for low temperatures and \cite{Colgan2008} for high temperatures, and the background heating is set to balance the initial cooling due to radiation (no thermal conduction in the initial state due to the isothermal condition), i.e.
\begin{equation}
    \mathcal{H} = \rho_0\, \Lambda(T_0),
\end{equation}
such that the heating is constant and uniform, and the initial configuration is in thermal equilibrium. Note though that the form of the heating is known to play a non-trivial role in the condensation process \citep{Brughmans2022}. Exploration of different heating prescriptions might be interesting to investigate in our model in the future. This will lead us to a better understanding about the coronal condensation mechanism in a reconnecting current sheet.

\section{Results}\label{sec:results}
Now, we investigate the stability properties of the force-free coronal current sheet introduced in Sec. \ref{sec:config}. In Sec. \ref{sec:legolas}, we apply the \legolas{} code to chart its linear regime. Subsequently, we focus on the non-linear regime in Sec. \ref{sec:amrvac} using simulations.

\subsection{Linear regime}\label{sec:legolas}
In this section, the linear stability properties of our coronal current sheet are probed with the \legolas{} code \citep{Claes2020b}. Since this equilibrium's $B_{y0}$- and $B_{z0}$-gradients are strongly localised, we discretise the domain $x\in [-10, 10]$ unevenly, accumulating grid points near the centre using the algorithm and Gaussian profile described in \cite{DeJonghe2024}, with parameters $p_1 = 0.2$, $p_2 = 0$, $p_3 = 0.001$, and $p_4 = 5$, yielding a resolution of $757$ grid points. A perfectly conducting wall boundary condition is applied on either $x$-boundary. To obtain the full spectrum and accompanying eigenfunctions, we use \legolas{}'s \texttt{QR-cholesky} solver. However, to execute parameter sweeps efficiently, we apply the \texttt{inverse-iteration} solver to the eigenvalue problem.

\subsubsection{Spectrum and perturbations}\label{sec:spectrum}
\begin{figure}
    \centering
    \includegraphics[width=\linewidth]{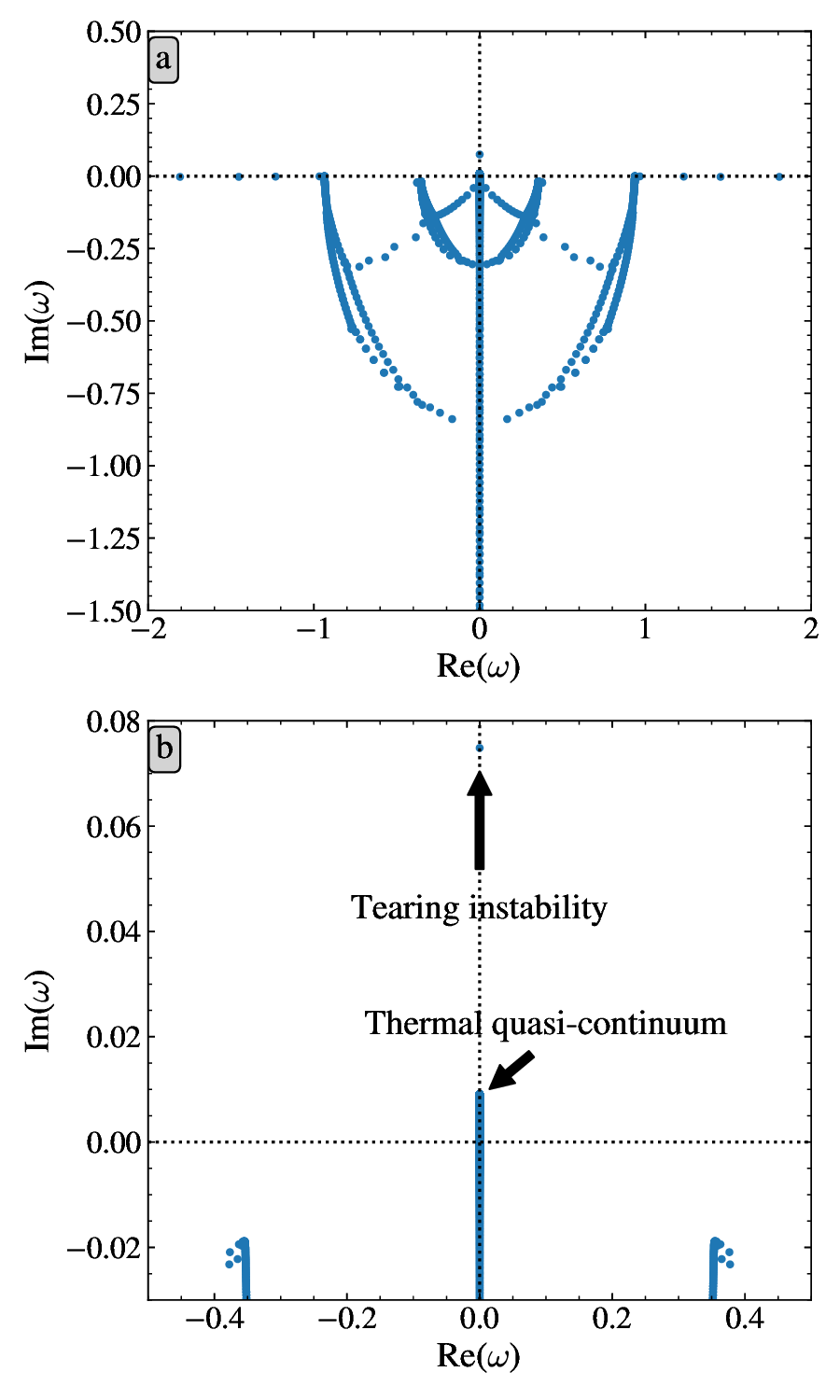}
    \caption{(a) Central region of the \legolas{} spectrum of complex eigenfrequencies for $k = 2\pi/15$. (b) Zoomed view of the region around the origin showing one discrete tearing instability and an unstable thermal quasi-continuum.}
    \label{fig:spectrum}
\end{figure}

For a wave vector $k_2 = 2\pi/15 \simeq 0.42$, $k_3 = 0$ (wavelength $\lambda = 150\,\mathrm{Mm}$), part of the eigenfrequency spectrum is shown in Fig. \ref{fig:spectrum}a. As expected in a resistive medium, the slow (inner semi-circle), Alfv\'en (outer semi-circle), and fast modes (outgoing sequences to the left and right) are all damped. Zooming in, Fig. \ref{fig:spectrum}b reveals one discrete tearing instability, as expected due to the magnetic shear, as well as an unstable thermal quasi-continuum. Similarly to the well-known slow and Alfv\'en continua, the combination of the non-adiabatic terms for radiative losses, heating, and parallel thermal conduction results in a continuous range of frequencies with singular eigenfunctions known as the thermal continuum \citep{VanDerLinden1991b, VanDerLinden1991c}. Note that since \legolas{} outputs a discrete amount of modes, a continuum appears as a sequence of modes, whilst in actuality every frequency along the imaginary axis between the origin and the most unstable mode in this sequence is an unstable eigenfrequency of the system. The continuum nature of these modes can be verified by the discontinuities in their perturbation profiles \citep[see e.g.][]{Goedbloed2019}. With the inclusion of resistivity though, the thermal continuum is replaced by a quasi-continuum, i.e. a frequency range densely packed with solutions \citep{Ireland1992}. In a quasi-continuum, the continuum modes' characteristic discontinuities are replaced with sharp, oscillatory behaviour.

\begin{figure}
    \centering
    \includegraphics[width=\linewidth]{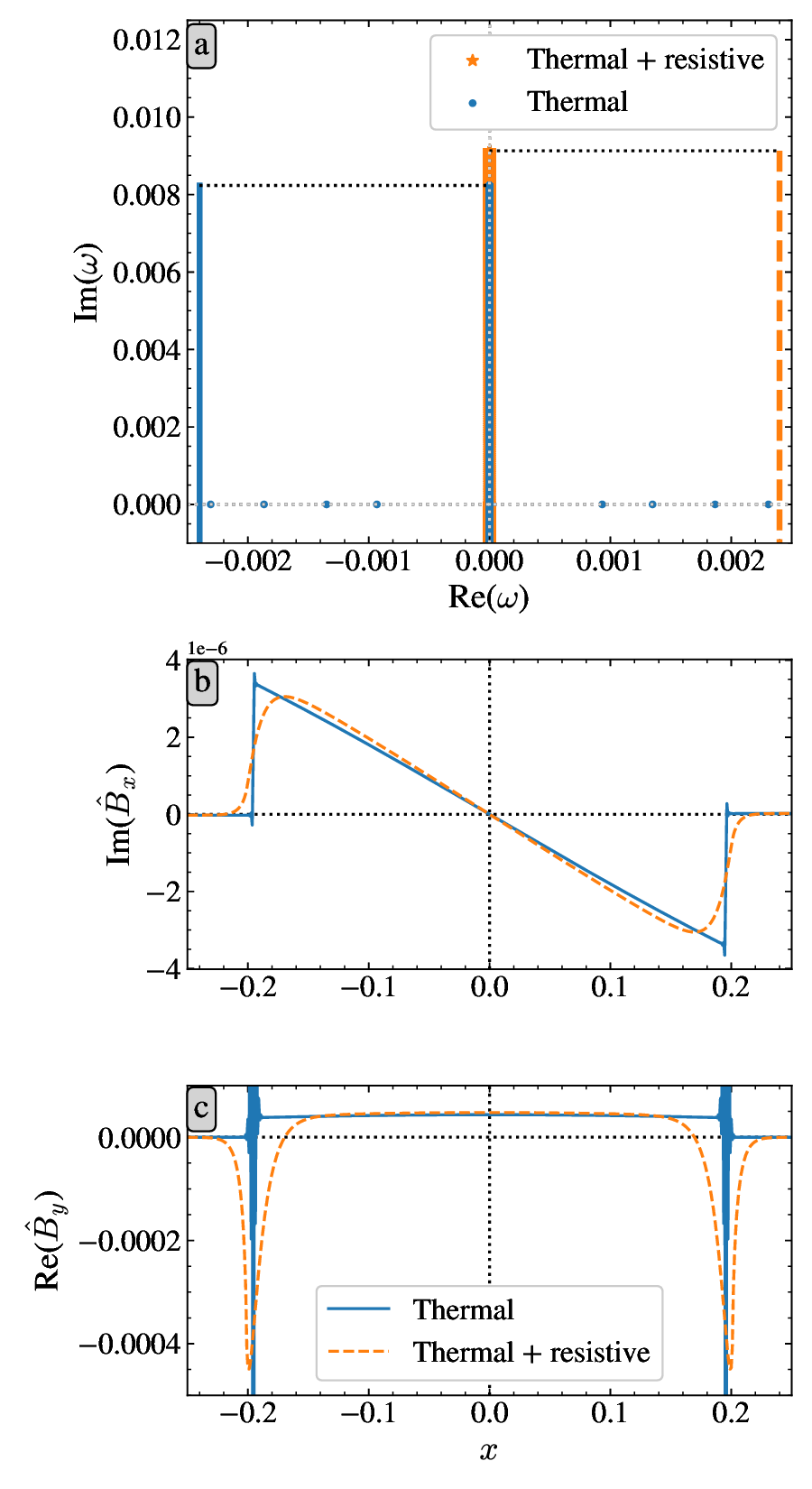}
    \caption{(a) Part of the \legolas{} eigenfrequency spectrum for $k = 2\pi/15$ comparing the non-resistive, thermal continuum (blue dots) and thermoresistive quasi-continuum (orange stars). The analytic prediction for each (quasi-)continuum range is shown in the corresponding colour on either side of the panel. (b) Comparison of the perturbation amplitudes of the magnetic field components $\hat{B}_x$ and (c) $\hat{B}_y$ for the thermal, non-resistive continuum mode $\omega \simeq 7.316\times 10^{-3}\,\im$ and thermoresistive quasi-continuum mode $\omega \simeq 7.319\times 10^{-3}\,\im$, highlighting the smoothing of the continuum mode's cd singularities by resistivity.}
    \label{fig:comparison}
\end{figure}

For a non-resistive, thermal setup (thermal conduction, radiative cooling, and background heating, but \emph{zero} resistivity), the part of the \legolas{} spectrum featuring the discretised continuum is shown with blue dots in Fig. \ref{fig:comparison}a. Taking the eigenfrequency $\omega \simeq 7.316\times 10^{-3}\,\im$ from this spectrum as an example, its $\hat{B}_x$ (purely imaginary) and $\hat{B}_y$ (purely real) perturbation amplitudes, also computed with \legolas{}, are shown as solid, blue lines in Figs. \ref{fig:comparison}b and \ref{fig:comparison}c, respectively. Note that due to the numerical approach, the discontinuities in the eigenfunctions are not well-resolved and instead feature sharp jumps around it, as is to be expected. In our case here, the current sheet configuration results in two locations of discontinuity, symmetric around the magnetic nullplane at $x = 0$, where $\bfk\cdot\bfb_0 = 0$.

\begin{figure}
    \centering
    \includegraphics[width=\linewidth]{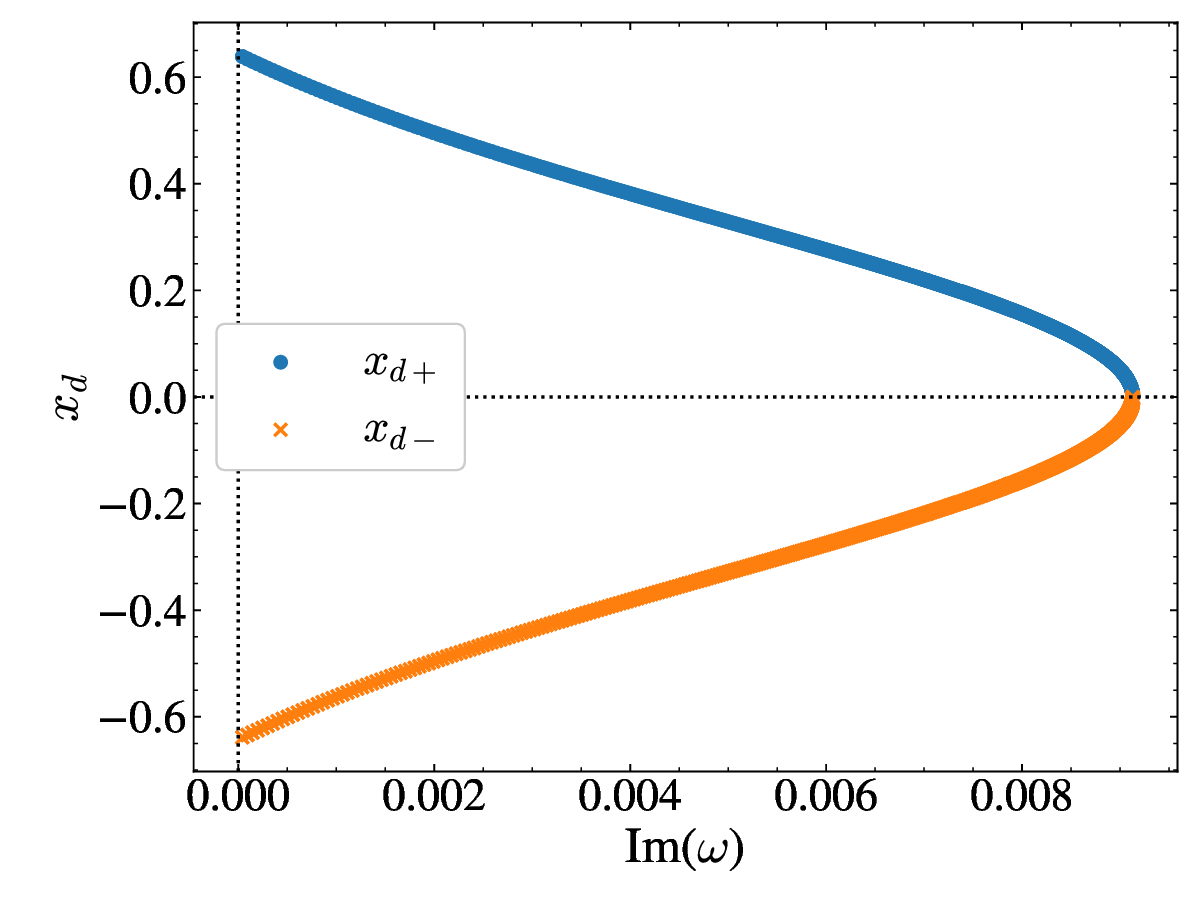}
    \caption{Locations of discontinuity in the eigenfunctions of thermal modes throughout the unstable thermal quasi-continuum of Fig. \ref{fig:spectrum} (note: $\mathrm{Re}(\omega) = 0$).}
    \label{fig:discontinuity}
\end{figure}

However, when we now compare the `thermoresistive' spectrum (thermal conduction, radiative cooling, background heating, \emph{and} resistivity), shown as orange stars in Fig. \ref{fig:comparison}a, to the thermal spectrum (blue dots), we observe that the inclusion of resistivity further extends the now quasi-continuum's range into the unstable half-plane. Comparing both ranges to their analytic predictions by \citet{VanDerLinden1991c} and \citet{Ireland1992} for the non-resistive and thermoresistive cases, respectively (shown on the figure axes), we find perfect agreement in both cases. In the overlapping area of the thermal and thermoresistive continua, the resistive layer smooths out the discontinuities in the eigenfunctions compared to the non-resistive, thermal continuum modes. To illustrate this, the $\hat{B}_x$ and $\hat{B}_y$ perturbation amplitudes (where this modification is most profound) of \legolas{}'s thermoresistive eigenfrequency $\omega \simeq 7.319\times 10^{-3}\,\im$ (the closest computed thermoresistive mode to \legolas{}'s thermal $\omega \simeq 7.316\times 10^{-3}\,\im$) are drawn as dashed orange lines in Figs. \ref{fig:comparison}b and \ref{fig:comparison}c, respectively. Note that at the most unstable end of the thermoresistive quasi-continuum, the eigenfunctions feature only one `lifted discontinuity', namely at $x = 0$. As you move away from the most unstable end towards the stable half-plane (downwards in Fig. \ref{fig:comparison}a), the lifted discontinuities manifest on either side of the domain's centre ($x=0$), and thus magnetic nullplane, at $x = x_{\mathrm{d}-} < 0$ and $x = x_{\mathrm{d}+} > 0$. As shown in Fig. \ref{fig:discontinuity}, we find that $x_{\mathrm{d}+} = -x_{\mathrm{d}-} \equiv x_{\mathrm{d}}$, and $x_\mathrm{d}$ increases up to $x_\mathrm{d} \sim 0.64$ for decreasing $\mathrm{Im}(\omega)$ before reaching the stable point.

\begin{figure}
    \centering
    \includegraphics[width=\linewidth]{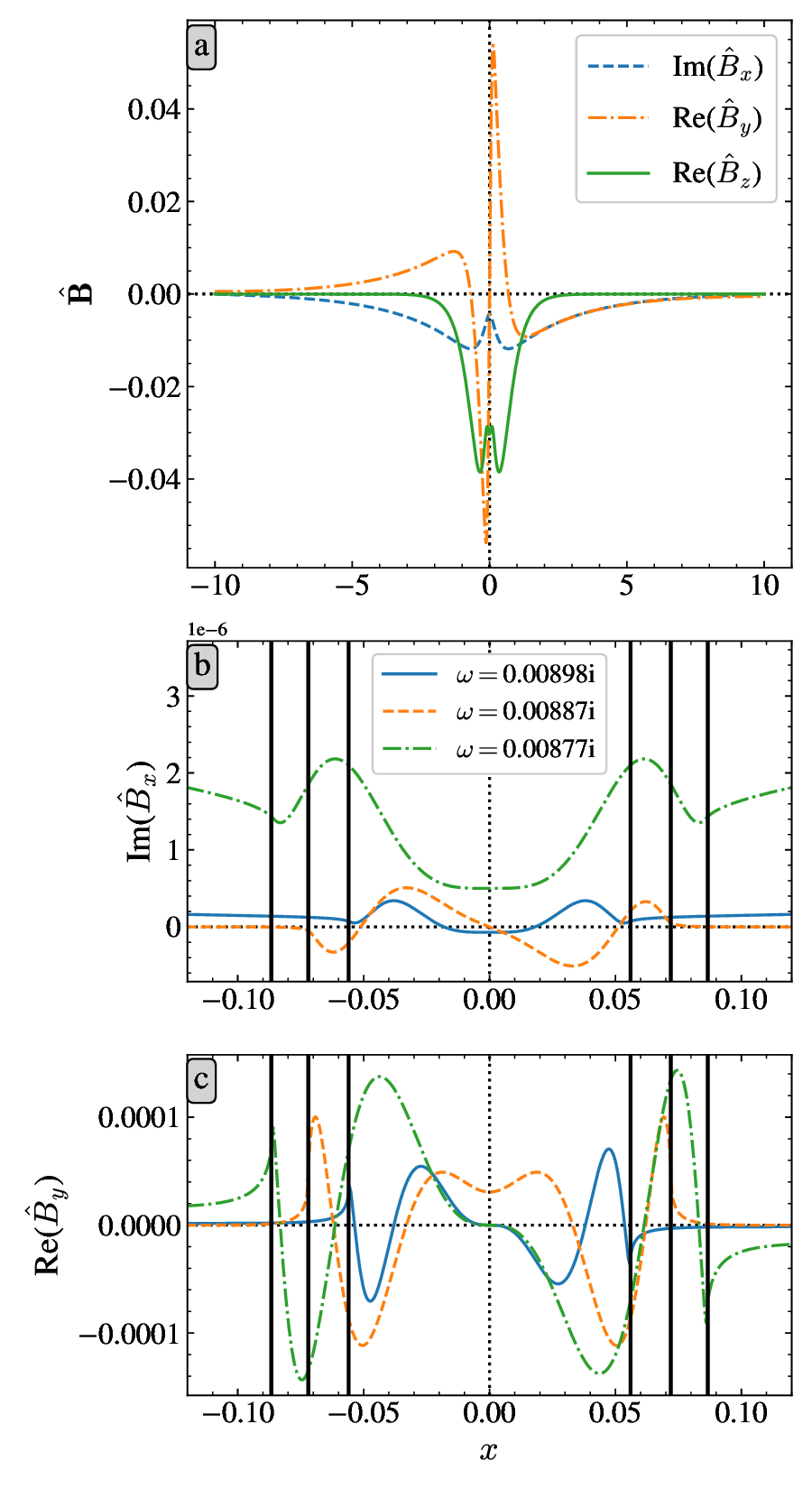}
    \caption{(a) Non-zero magnetic field perturbations of the tearing instability. (b) Magnetic field component $\hat{B}_x$ and (c) $\hat{B}_y$ perturbation amplitudes of an arbitrary selection of quasi-continuum modes. The approximate locations of the transitions are indicated with solid, black lines.}
    \label{fig:ef}
\end{figure}

Before discussing the upper, non-overlapping part of the thermoresistive quasi-continuum, we first turn to the tearing instability. Its magnetic field perturbation amplitude parallel to the direction of the equilibrium's polarity inversion, $\hat{B}_y$, is zero at the magnetic nullplane ($x=0$) with extrema of opposite sign on either side. Simultaneously, the magnetic perturbation amplitude in the direction of the equilibrium's variation, $\hat{B}_x$, features symmetric extrema on either side of the nullplane, as expected of a tearing mode. This can be seen in Fig. \ref{fig:ef}a alongside the third magnetic field component, $\hat{B}_z$.

Now, as it turns out, the behaviour of the modes in the upper, non-overlapping part of the thermoresistive quasi-continuum is different between the two discontinuities compared to the outside regions. Since the resistivity smooths out the discontinuity in the magnetic field perturbation, the inner and outer solutions of the $\bfb$-perturbation have to be matched at the discontinuity locations. This results in two types of modes, here referred to as Type I and Type II modes. Type I modes have magnetic perturbation amplitudes that have the same shape as the tearing mode outside of the thin layer between their perturbation discontinuities.  To match the symmetry (antisymmetry) of the tearing mode's $\hat{B}_x$ ($\hat{B}_y$) profile, their oscillatory part in $\hat{B}_x$ ($\hat{B}_y$) between their discontinuities is also symmetric (antisymmetric). The $\hat{B}_x$ and $\hat{B}_y$ amplitudes are shown for two such modes with solid, blue and dash-dotted, green lines in Figs. \ref{fig:ef}b and \ref{fig:ef}c, respectively. The approximate locations of transition (lifted discontinuities) between inner and outer solution are indicated with solid, black lines. Type II modes on the other hand have an antisymmetric (symmetric) oscillatory part in $\hat{B}_x$ ($\hat{B}_y$) between their discontinuities, and thus cannot match with a tearing-like outer solution. An example of a Type II mode's magnetic perturbation amplitudes is drawn in dashed, orange lines in Figs. \ref{fig:ef}b and \ref{fig:ef}c. Note though that a Type II $\hat{B}_x$ ($\hat{B}_y$) perturbation behaves like a Type I $\hat{B}_y$ ($\hat{B}_x$) perturbation and vice versa.

\subsubsection{Dominant wave vector}\label{sec:lin-dom}
\begin{figure}
    \centering
    \includegraphics[width=\linewidth]{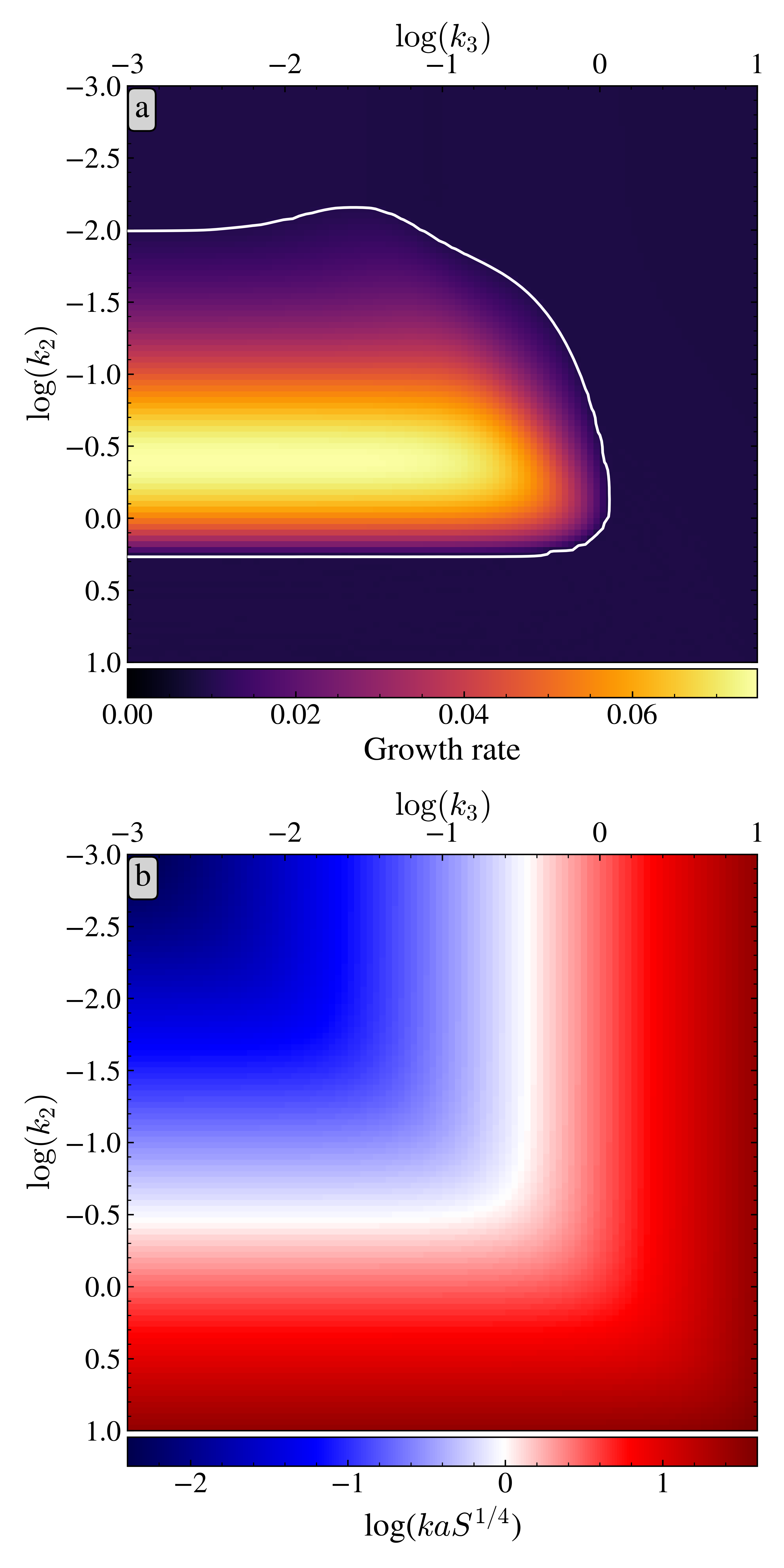}
    \caption{Growth rate of the most unstable mode in each spectrum for combinations of $k_2$ and $k_3$, for the thermoresistive current sheet described in Sec. \ref{sec:config}. The white contour corresponds to $\imag(\omega) = 0.01$ for the resistive, non-thermal current sheet. (b) $\log(kaS^{1/4})$ as a measure of constant-$\psi$ ($kaS^{1/4} \gg 1$) versus nonconstant-$\psi$ ($kaS^{1/4} \ll 1$) tearing regimes.}
    \label{fig:kvar}
\end{figure}

Like many unstable equilibria, the current sheet considered here is not equally unstable to all wavelengths. Therefore, we perform a parameter sweep varying $k_2$ and $k_3$ simultaneously with $100\times 100$ combinations, evaluated with \legolas{}. The maximal growth rate is visualised for each wave vector in Fig. \ref{fig:kvar}a for the thermoresistive current sheet described in Sec. \ref{sec:config}. The tearing instability dominates at intermediate values of $k_2$, remaining approximately constant for increasing $k_3$ until a drop-off occurs around $k_3 \sim 1$. In the resistive, non-thermal equivalent (no thermal conduction, radiative losses, or background heating), the current sheet is found to be stable outside of this strongly tearing-dominated region of the $(k_2, k_3)$ parameter space, indicated by the non-thermal $\imag(\omega) = 0.01$ contour in white in Fig. \ref{fig:kvar}a, outside of which the tearing growth rate rapidly drops to zero. In the thermoresistive case shown here, the tearing growth rate vanishes, or is smaller than the thermal quasi-continuum's maximal growth rate, for the $(k_2, k_3)$ combinations outside of the contour, and thus dominated by thermal instability at these wavelengths.

\begin{figure*}
    \centering
    \includegraphics[width=0.9\linewidth]{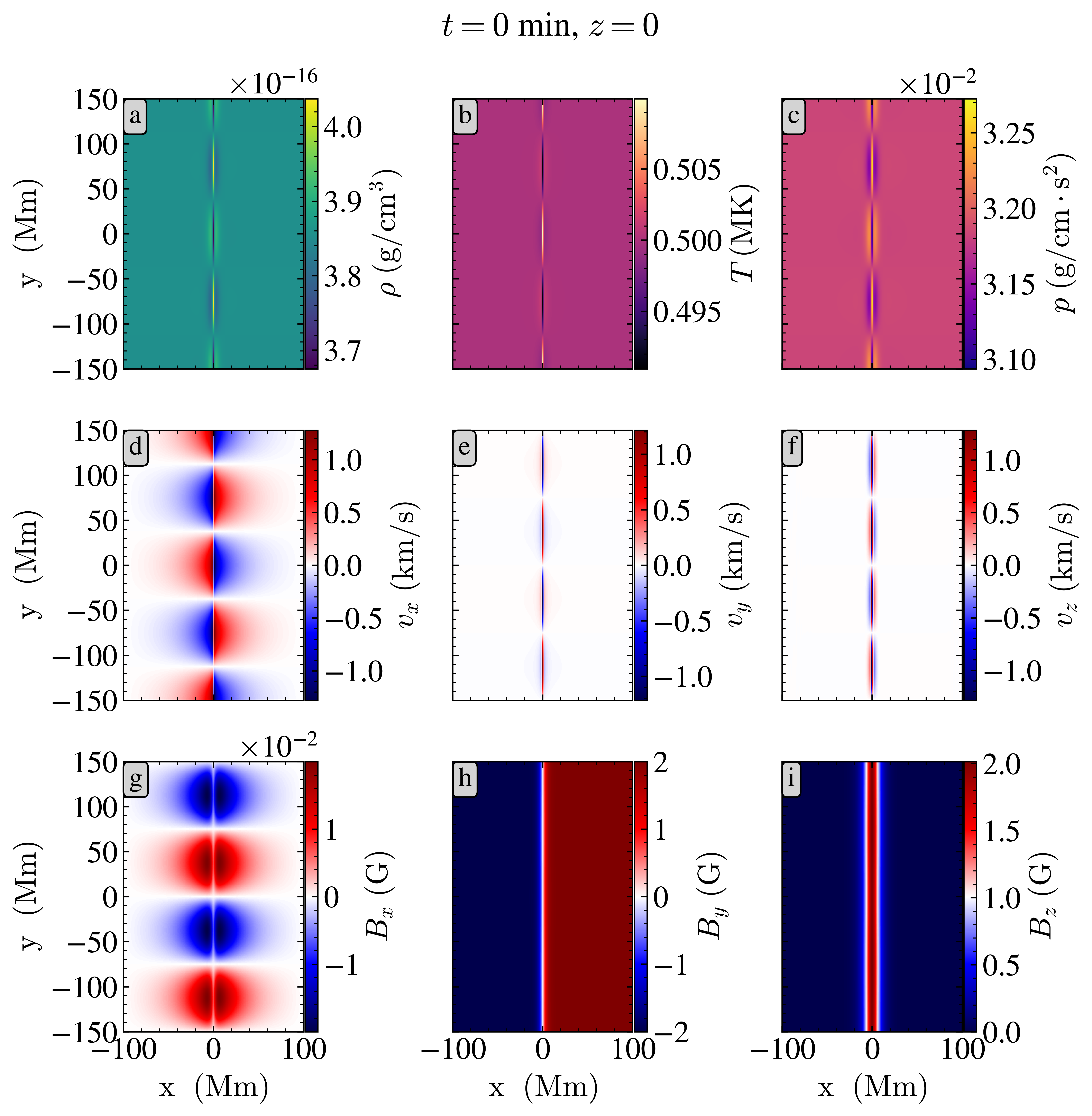}
    \caption{Initial state of the (a) density, (b) temperature, (c) thermal pressure, (d) $v_x$, (e) $v_y$, and (f) $v_z$ velocity components, and (g) $B_x$, (h) $B_y$, and (i) $B_z$ magnetic field components.}
    \label{fig:initial}
\end{figure*}

The most unstable wave vector is found to align with the axis corresponding to the reversing magnetic field component, i.e. the $y$-axis in our case, and its wave number is close to $k_2 \simeq 0.4$ (and thus to the wave number selected in Sec. \ref{sec:spectrum}) with a growth rate of $\imag(\omega_\mathrm{t}) \simeq 7.48\times 10^{-2}$ (or $9.61\times 10^{-4}\ \mathrm{rad}\,\mathrm{s}^{-1}$ in physical units). The decrease in growth rate is minimal when a small $k_3$-component is added though. Quantifying $kaS^{1/4}$ in Fig. \ref{fig:kvar}b as a means to distinguish between the constant-$\psi$ ($kaS^{1/4} \gg 1$, red region) and nonconstant-$\psi$ ($kaS^{1/4} \ll 1$, blue region) regimes in analytic theory reveals that the most unstable tearing modes (both parallel and oblique) lie at or close to the transition between both regimes, contrary to the results of \citet{Baalrud2012}.

In comparison to the tearing instability, the growth rate $\imag(\omega_\mathrm{c})$ of the unstable edge of the thermal quasi-continuum is more or less constant throughout the parameter space, and slightly less than an order of magnitude smaller than the maximal tearing growth rate, at $\imag(\omega_\mathrm{c}) \simeq 9.13\times 10^{-3}$ (or $1.17\times 10^{-4}\ \mathrm{rad}\,\mathrm{s}^{-1}$ in physical units).

\subsection{Non-linear regime}\label{sec:amrvac}

To investigate this thermoresistive Harris sheet's non-linear evolution, we set it up with \amrvac{} \citep{Porth2014, Keppens2021, Keppens2023} in a spatial domain ranging from $-10$ to $10$ in the $x$-direction ($200\ \mathrm{Mm}$), and $-15$ to $15$ in the $y$- and $z$-directions ($300\ \mathrm{Mm}$), to be close to twice the most unstable wavelength identified in Sec. \ref{sec:lin-dom}. Initially, this domain is discretised with $256\times 384^2$ cubic cells (along the $x$-, $y$- and $z$-directions respectively), after which the code's adaptive mesh refinement (AMR) level is set to $2$, resulting in a maximum resolution of $512\times 768^2$. With our length unit of $\bar{L} = 10^9\,\mathrm{cm}$, this corresponds to a physical resolution of $\sim 391\,\mathrm{km}$ in each direction. To perturb the Harris equilibrium, we add the perturbations of the $k_2 \simeq 0.42$ tearing instability (the most unstable wave number identified in Sec. \ref{sec:lin-dom}) in all variables, for $t = 0$ in equation (\ref{eq:fourier}) and normalised such that the maximal $B_y$-perturbation is $1$ per cent of $B_0$, i.e. $\max (B_{y1}) = 0.01\,\max (B_{y0})$. The resulting initial setup is visualised at $z = 0$ in Fig. \ref{fig:initial} (note that it looks identical for all $z$ because $k_3 = 0$). In line with \legolas{}'s boundary conditions, we apply perfectly conducting wall boundaries in the $x$-direction and periodic boundaries in the $y$- and $z-$directions. Past the initial setup, the system is allowed to evolve in time according to equations (\ref{eq:vac-continuity})-(\ref{eq:vac-Bdivfree}), using a three-step Total Variation Diminishing Lax-Friedrichs (\textsf{tvdlf}) flux scheme, \textsf{minmod} limiter, and Courant step of $0.8$. The simulation was allowed to evolve for $129.7\ \mathrm{min}$, saving data with a temporal cadence of $77.8\ \mathrm{s}$ for a total of $101$ snapshots. This took $\sim 57\ \mathrm{hours}$ to complete on $512$ cores.

\begin{figure*}
    \centering
    \includegraphics[width=0.9\linewidth]{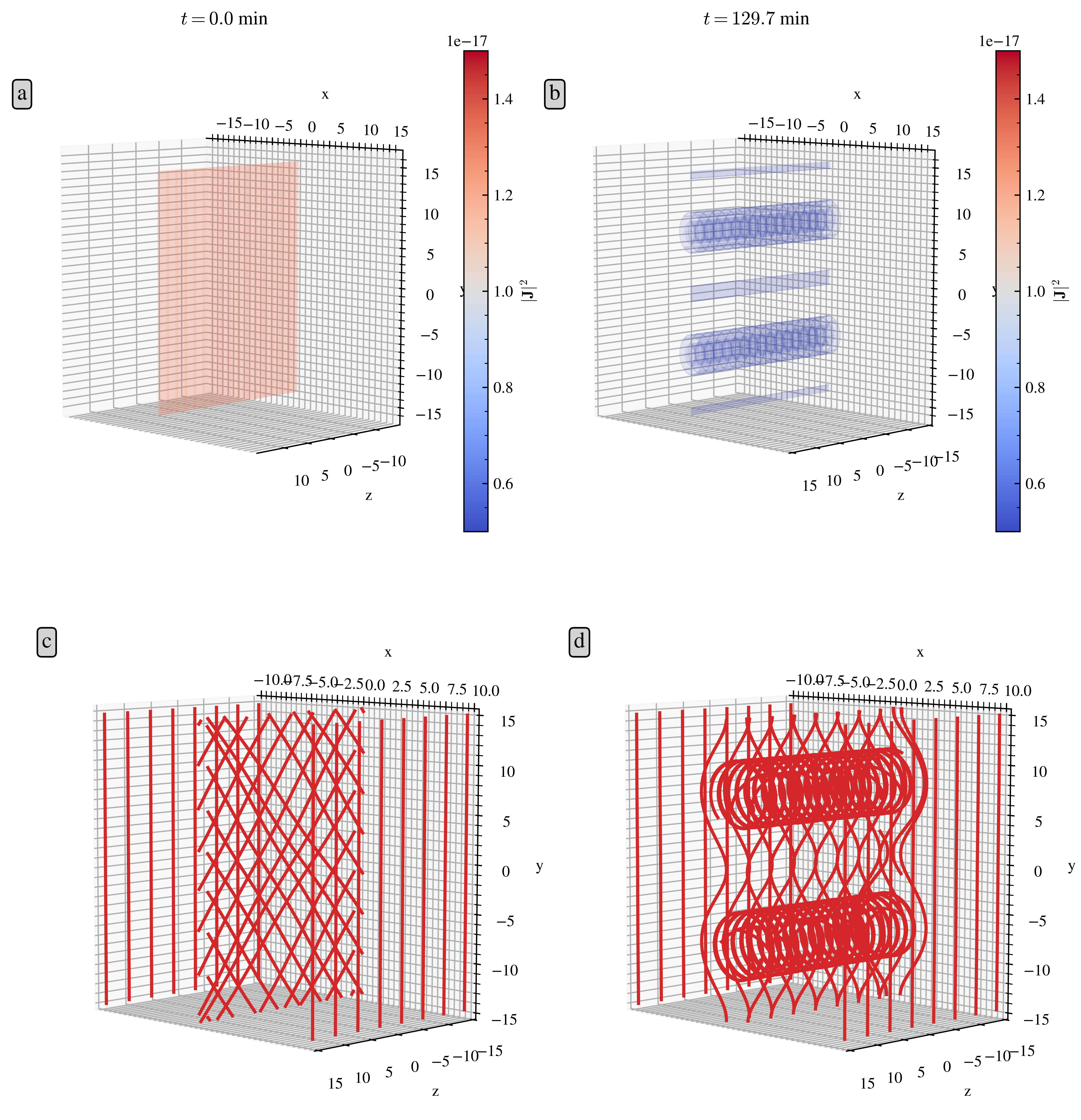}
    \caption{(a-b) Isosurfaces of the current density squared $\bfj^2$ at the (a) beginning and (b) end of the simulation. (c-d) Selection of magnetic field lines at the (c) beginning and (d) end of the simulation.}
    \label{fig:isosurface-J2}
\end{figure*}

First, we generally describe the observed evolution. At the initial time, the setup deviates only slightly from the Harris equilibrium. The current is mostly confined close to the current sheet centre at $x=0$, with near-planar isosurfaces parallel to the $yz$-plane, and the magnetic field almost parallel to the $y$-axis, except near the current sheet centre at $x = 0$, where it reverses its direction rapidly across the current sheet. This is illustrated in Figs. \ref{fig:isosurface-J2}a and \ref{fig:isosurface-J2}c. At the end of the simulation, after $129.7\,\mathrm{min}$, the sheet inside the simulation domain is torn up into two plasmoids, whose helical flux rope structures are clearly visible in Fig. \ref{fig:isosurface-J2}d, and the corresponding current density distribution is shown as isosurfaces of $J^2$ in Fig. \ref{fig:isosurface-J2}b.  Considering we limited the domain to twice the most unstable wavelength in the $y$-direction, this behaviour is in line with expectations.

\begin{figure}
    \centering
    \includegraphics[width=\linewidth]{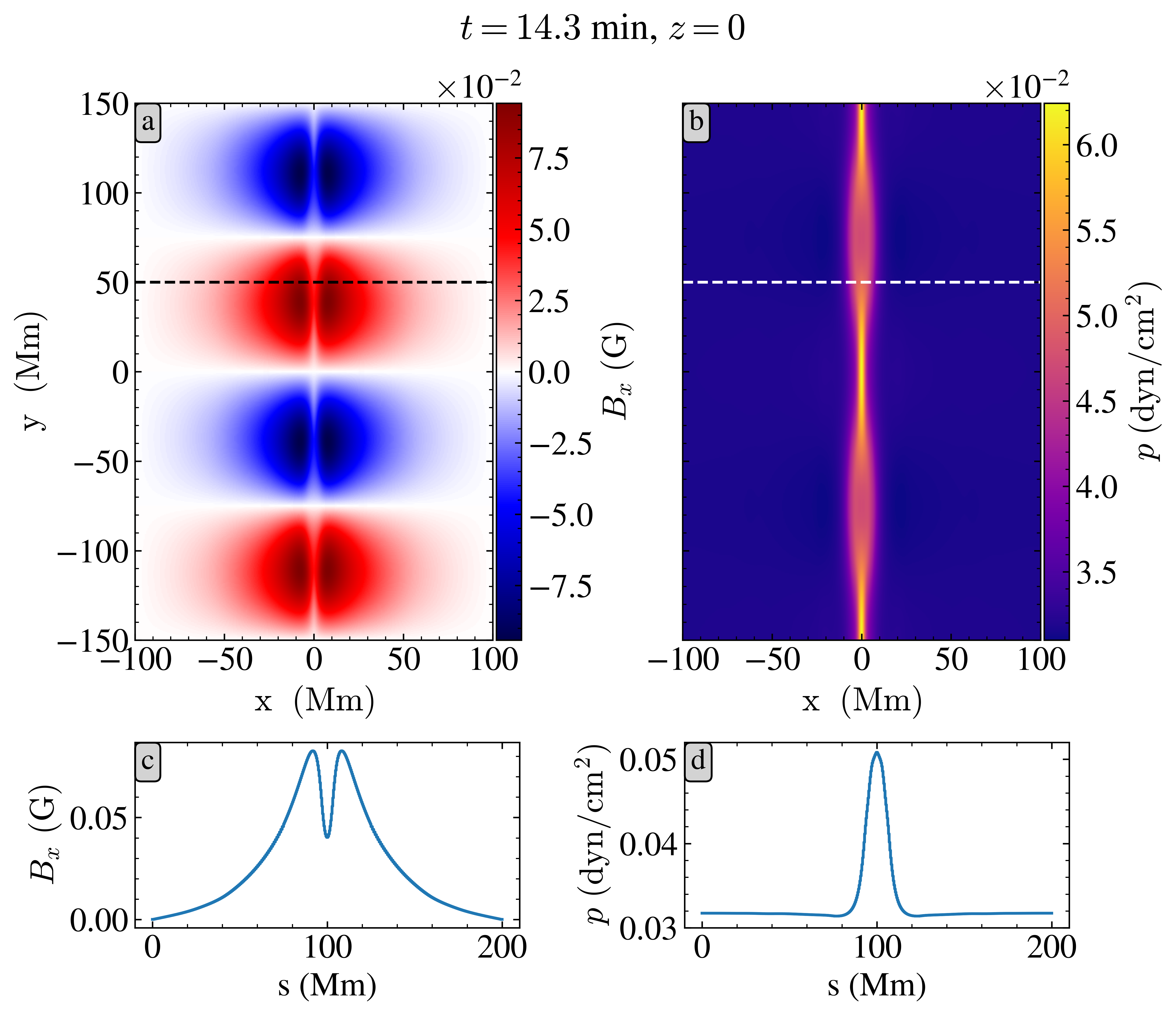}
    \caption{The (a) $B_x$ magnetic field component and (b) thermal pressure in the early stages of the non-linear evolution, at $t = 14.3\ \mathrm{min}$ and $z=0$. (c) $B_x$ and (d) $p$ with respect to path length $s$ along the dashed lines in panels (a) and (b).}
    \label{fig:early-snapshot}
\end{figure}

\begin{figure}
    \centering
    \includegraphics[width=\linewidth]{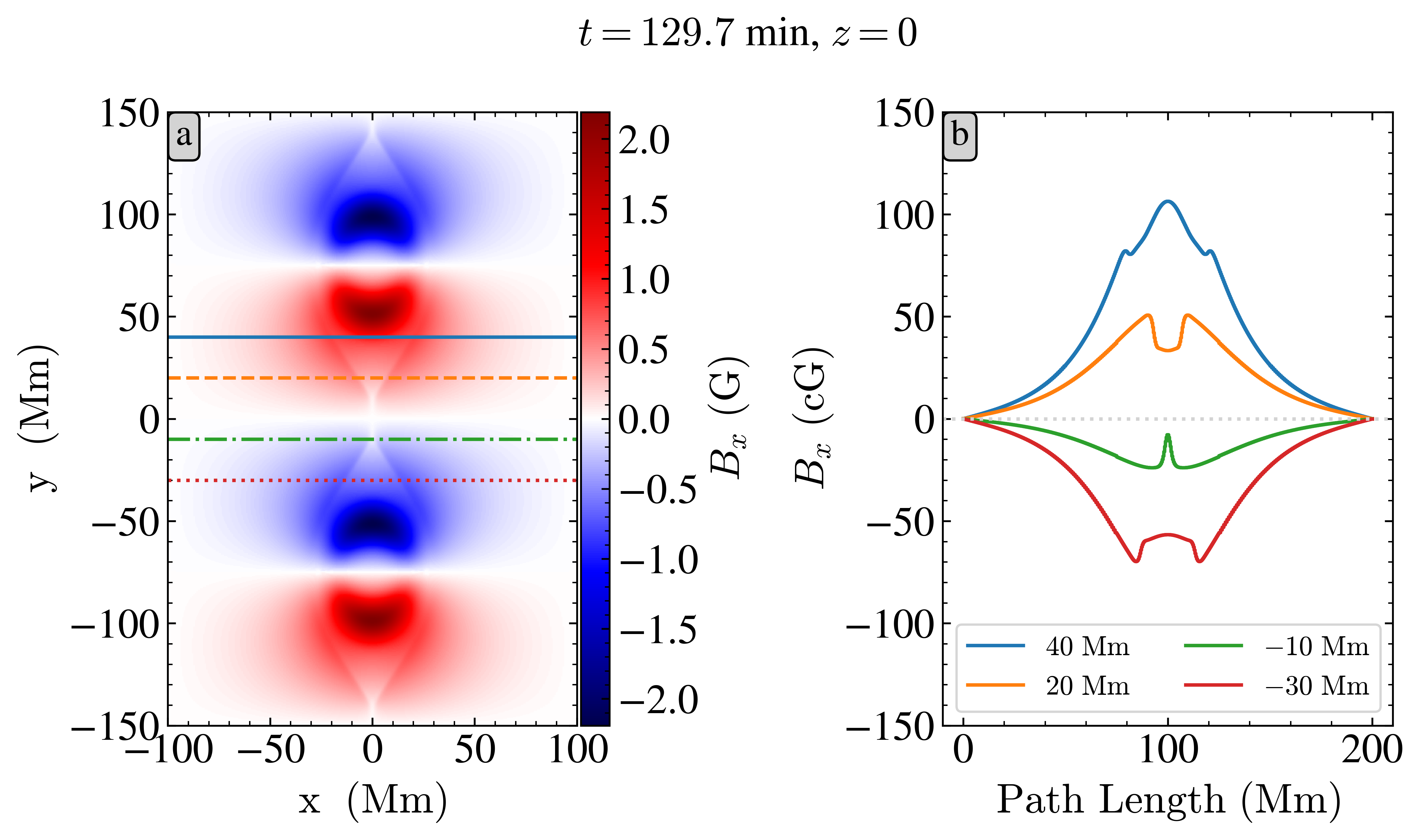}
    \caption{(a) The $B_x$ magnetic field component at $z=0$ at the end of the simulation, $t = 129.7\ \mathrm{min}$. (b) $B_x$ along each cut in panel (a), showing how the tearing signature changes within the plasmoid.}
    \label{fig:Bx-final}
\end{figure}

Due to the dominance of the tearing instability in the linear regime by an order of magnitude, the early evolution of the current sheet progresses as one would expect from tearing, with plasmoids forming and growing in size. Fig. \ref{fig:early-snapshot}a shows how after $14.3$ minutes, the magnetic field perturbation in the $x$-direction has quadrupled in size compared to the initial perturbation in Fig. \ref{fig:initial}g whilst maintaining a clean tearing signature, which is shown in Fig. \ref{fig:early-snapshot}c for the dashed line. Since we initialised with the dominant tearing instability, this is a sign of very little mode mixing with the thermal modes at this stage of the evolution. Notably, the tearing signature remains relatively clean in the outer regions throughout the simulation, but the central behaviour is modified, as shown in Fig. \ref{fig:Bx-final}. Now, comparing Fig. \ref{fig:early-snapshot}b to Fig. \ref{fig:initial}c, we see that the thermal pressure inside the current sheet has doubled since the initial time, whilst the environment's pressure has remained constant. Furthermore, we see that the thermal pressure inside the plasmoids is larger than that of the surroundings, but smaller than the pressure in the sheet between the plasmoids. This is due to the conversion of magnetic into internal energy in the current sheet by Ohmic dissipation, and the conversion of magnetic to internal energy in the reconnection process.

\begin{figure}
    \centering
    \includegraphics[width=\linewidth]{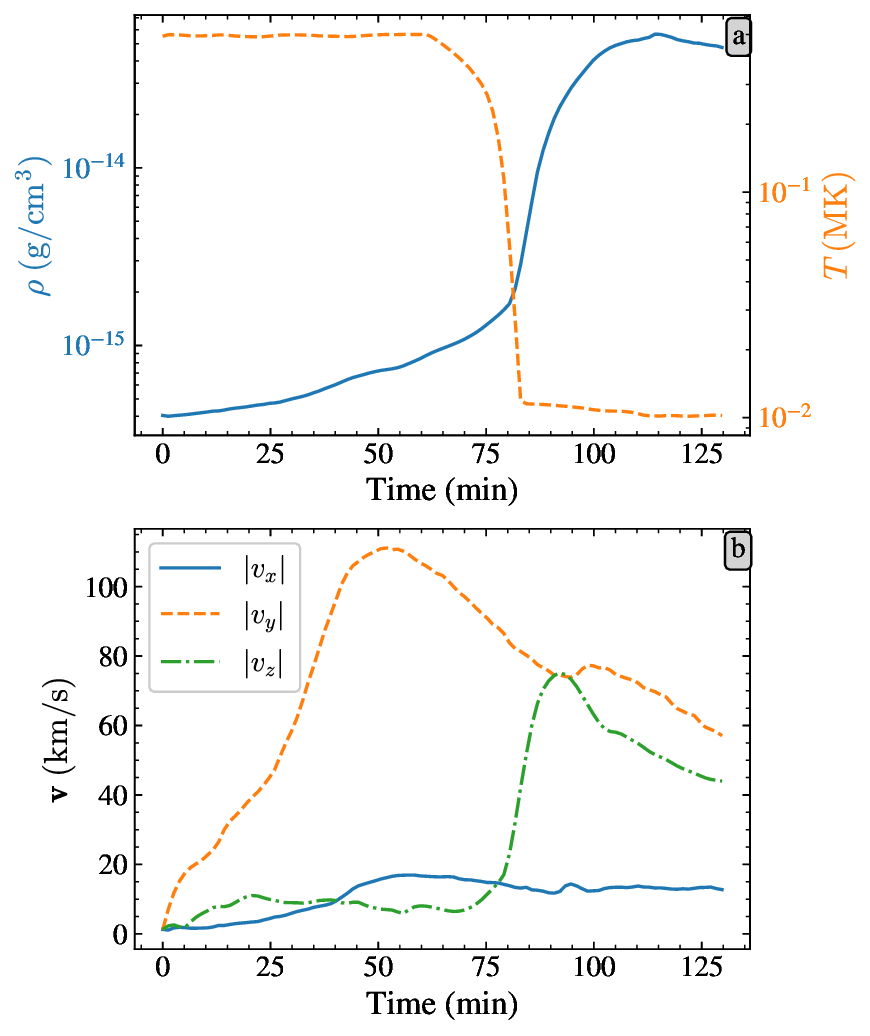}
    \caption{(a) Maximal density $\rho$ and minimum temperature $T$ in the simulation domain as a function of time. (b) Maximal value of each velocity component magnitude in the simulation domain as a function of time.}
    \label{fig:t-evol}
\end{figure}

\begin{figure}
    \centering
    \includegraphics[width=\linewidth]{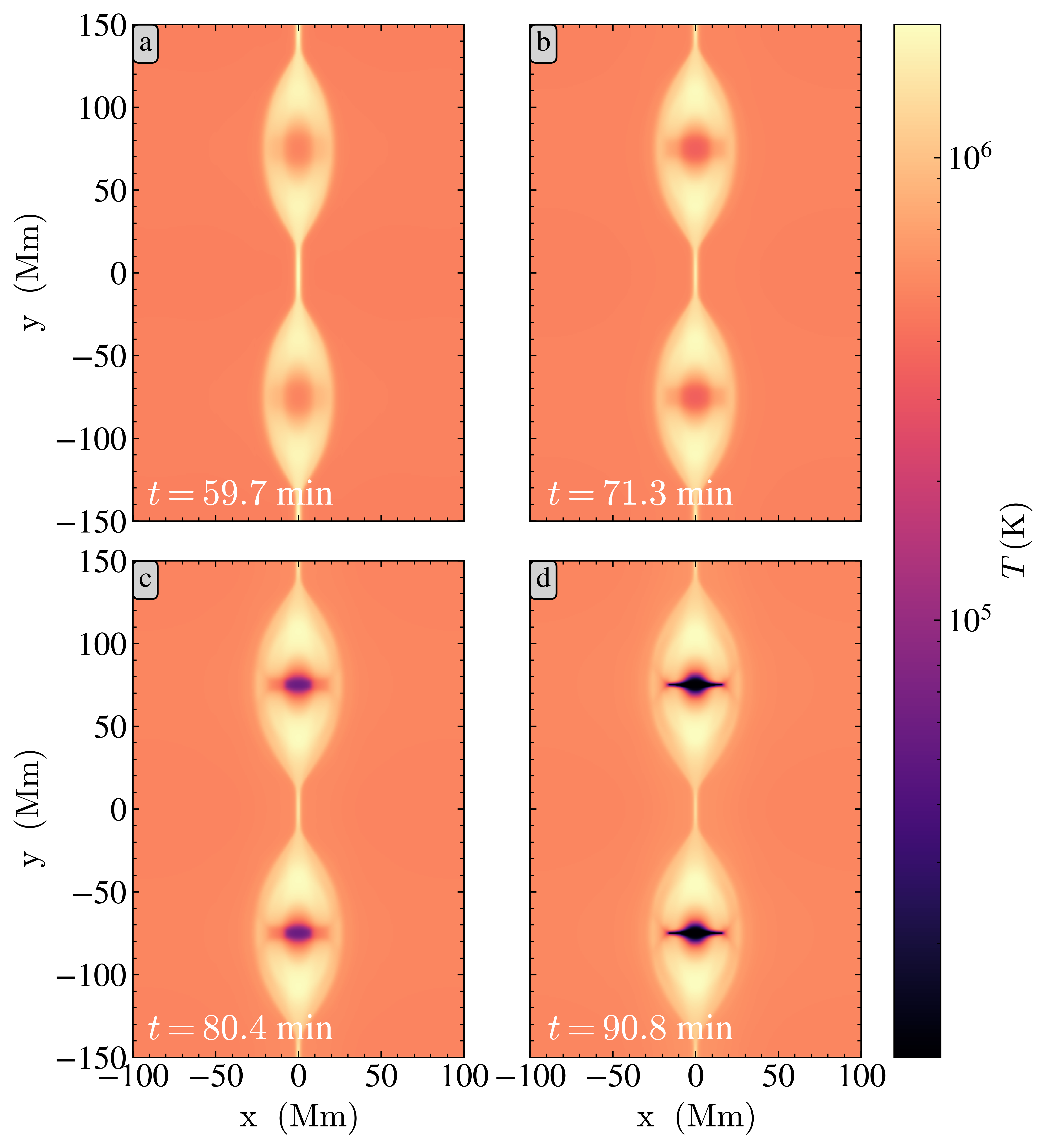}
    \caption{Temperature at the $z=0$ plane at (a) $t = 59.7\ \mathrm{min}$, (b) $t = 71.3\ \mathrm{min}$, (c) $t = 80.4\ \mathrm{min}$, and (d) $t = 90.8\ \mathrm{min}$.}
    \label{fig:temp-evol}
\end{figure}

The constant pressure away from the current sheet and the developing plasmoids, towards the $x$-boundaries, is maintained because in that region the radiative losses are balanced by the constant background heating. Inside the current sheet and growing plasmoids though, we expect an imbalance between these two effects with radiative cooling dominating, and consequently, that the density and temperature evolve towards a catastrophic stage, increasing density and cooling the plasma significantly in a self-reinforcing cycle. Therefore, we track the maximal density and minimal temperature in the simulation domain in Fig. \ref{fig:t-evol}a. Here, it becomes clear that initially the density inside the plasmoids is increasing steadily until around $t = 60\ \mathrm{min}$, at a (linearly-fitted) rate of $1.21\times 10^{-19}\ \mathrm{g}\,\mathrm{cm}^{-3}\,\mathrm{s}^{-1}$ for $t < 59.7\ \mathrm{min}$, after which time the density increases sharply by an order of magnitude at a rate of $2.29\times 10^{-17}\ \mathrm{g}\,\mathrm{cm}^{-3}\,\mathrm{s}^{-1}$ in the interval $79.1\ \mathrm{min} < t < 89.5\ \mathrm{min}$, paired with a drop of more than an order of magnitude in temperature at a rate of $-4.49\times 10^2\ \mathrm{K}\,\mathrm{s}^{-1}$ in the interval $75.2\ \mathrm{min} < t < 85.6\ \mathrm{min}$, reminiscent of prominence structures and coronal rain. As can be seen from the temperature evolution at $z = 0$ in Fig. \ref{fig:temp-evol}, the thermal collapse first occurs at the flux rope axis (Fig. \ref{fig:temp-evol}c) before spreading perpendicularly to the current sheet (Fig. \ref{fig:temp-evol}d), forming `wings'. Similarly to Fig. \ref{fig:t-evol}a, Fig. \ref{fig:t-evol}b shows how each velocity component's maximal magnitude changes. Most notably, coinciding with the drop in temperature, is a strong increase in the velocity component parallel to the flux rope axes, revealing a dynamical instability in the system at the condensation onset time. We will return to the discussion of the velocity profiles shortly.
    
\begin{figure}
    \centering
    \includegraphics[width=\linewidth]{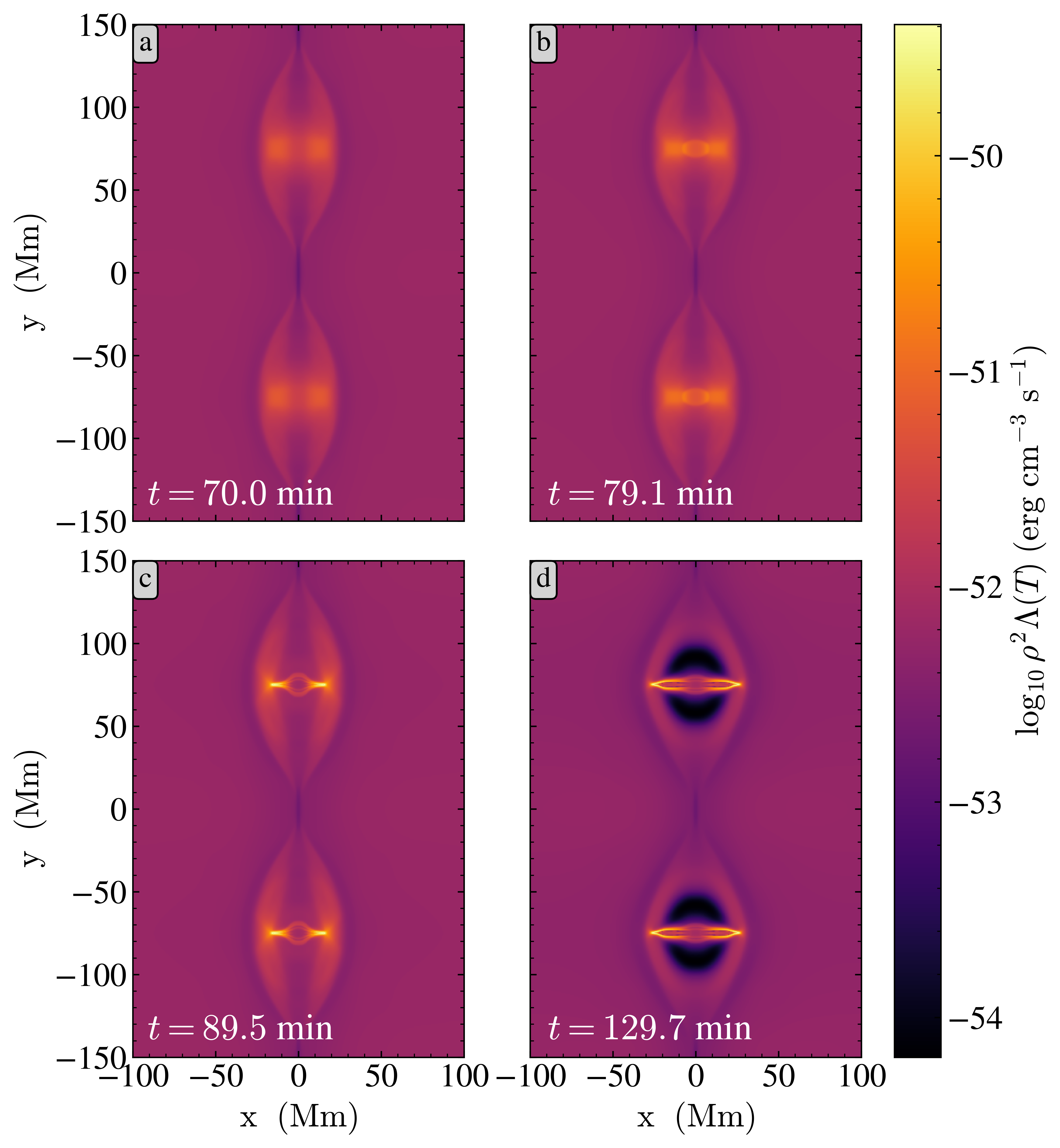}
    \caption{Radiative cooling $\rho^2\Lambda(T)$ in the $z = 0$ plane at (a) $t = 70\ \mathrm{min}$, (b) $t = 79.1\ \mathrm{min}$, (c) $t = 89.5\ \mathrm{min}$, and (d) $t = 129.7\ \mathrm{min}$.}
    \label{fig:cooling}
\end{figure}

To understand the condensation process in Fig. \ref{fig:temp-evol}, and the formation of these wings, it merits looking at the radiative losses. For this reason, $\rho^2\Lambda(T)$ is quantified throughout the condensation process in Fig. \ref{fig:cooling}.  Panels a, b, and c in this figure shows the radiative losses in the snapshots preceding the ones shown in panels b, c, and d of Fig. \ref{fig:temp-evol}, respectively. Fig. \ref{fig:cooling}d shows the radiative cooling at the end of the simulation. Initially, in Fig. \ref{fig:cooling}a, we observe the formation of regions with higher radiative losses on both sides of the flux rope axis inside each plasmoid, due to density enhancement there. These emissions result in a vaguely plus-shaped region of lower temperature inside the plasmoid, as seen in Fig. \ref{fig:temp-evol}b. Only a bit later, in Fig. \ref{fig:cooling}b, do we observe a strong emission zone at the flux rope axis, leading to the axial, catastrophic cooling in Fig. \ref{fig:temp-evol}c. Finally, the wings become the strongest radiative region in Fig. \ref{fig:cooling}c, expanding the dense plasma from the flux rope axis towards the plasmoid edges, perpendicularly to the current sheet.

\begin{figure*}
    \centering
    \includegraphics[width=0.9\linewidth]{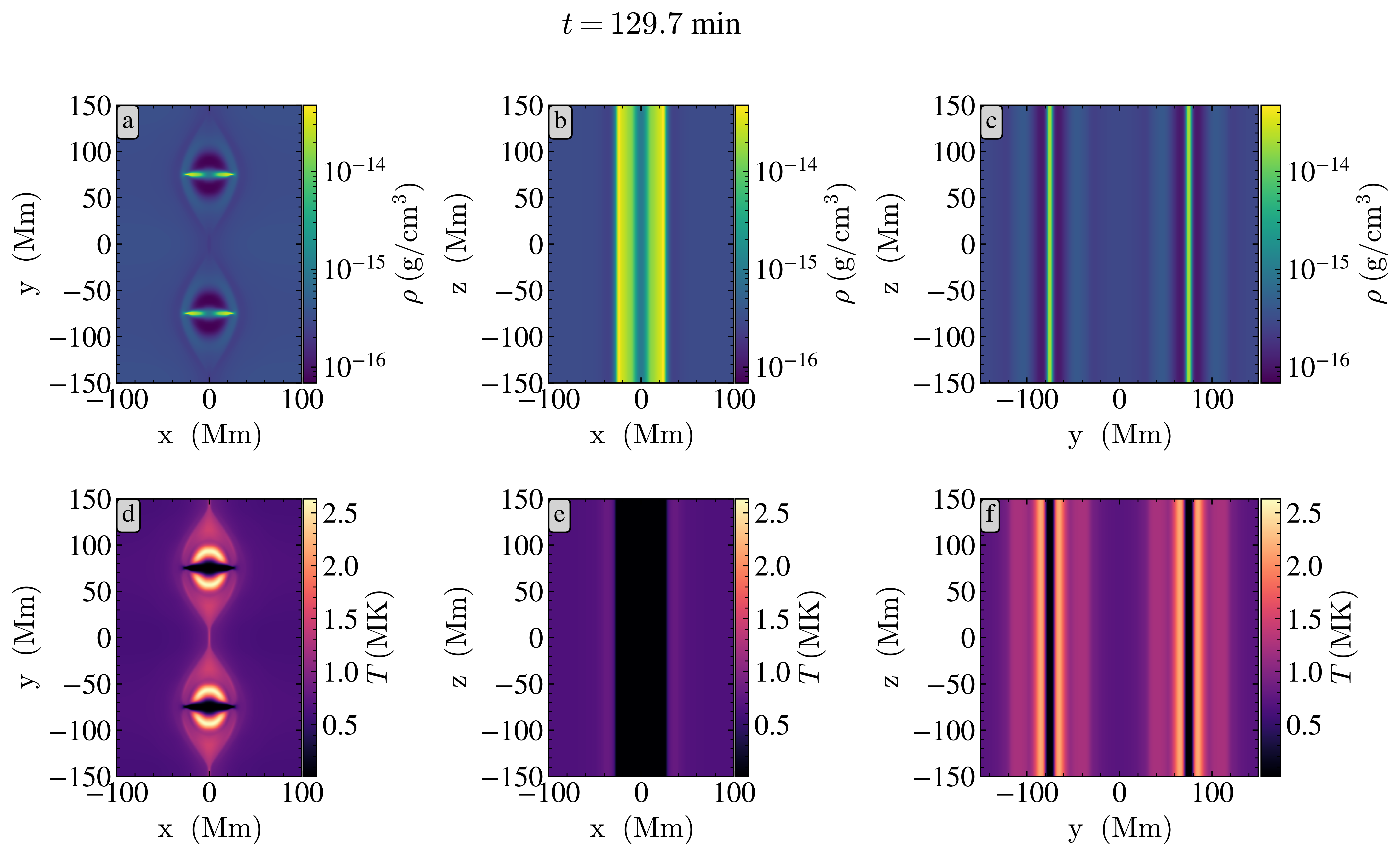}
    \caption{(a-c) Density and (d-f) temperature at $t = 129.7\ \mathrm{min}$ in the (left to right) $z = 0\ \mathrm{Mm}$, $y = 75\ \mathrm{Mm}$, and $x = 15\ \mathrm{Mm}$ planes.}
    \label{fig:rhoT-end}
\end{figure*}

At the end of the simulation, as shown in Fig. \ref{fig:rhoT-end}d, the condensations are flanked on both sides by a hot, weakly radiative region. On the other hand, Fig. \ref{fig:cooling}d shows that the strongest radiative losses occur at the edges of the condensation's wings whilst the flux rope axis and inside of the wings are similar in radiative losses to the environment, and thus balanced by the background heating. Consequently, the temperature is not expected to drop much further, but the condensation region might expand slightly if the simulation were allowed to continue.

\begin{figure*}
    \centering
    \includegraphics[width=0.9\linewidth]{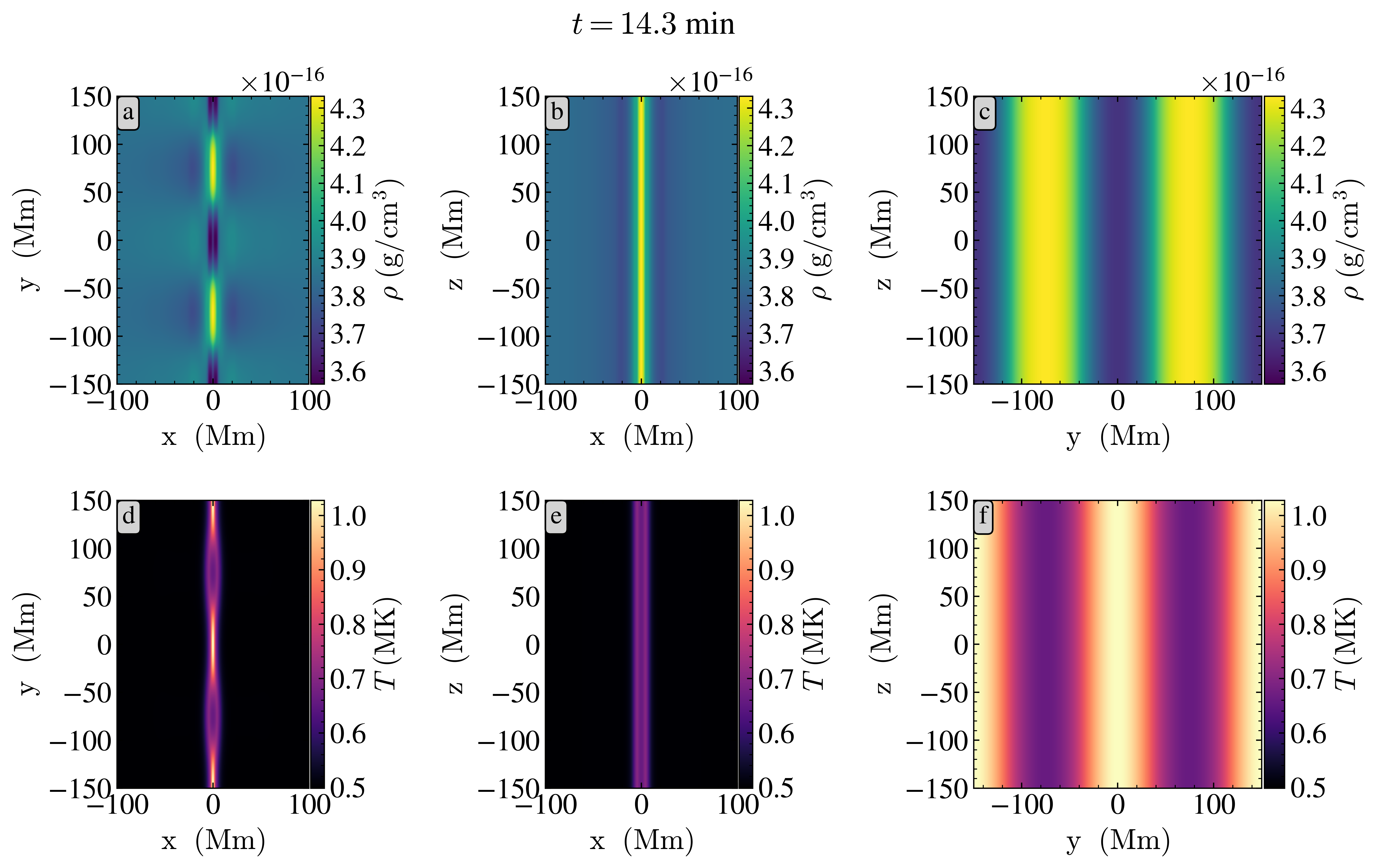}
    \caption{(a-c) Density and (d-f) temperature at $t = 14.3\ \mathrm{min}$ in the (left to right) $z = 0\ \mathrm{Mm}$, $y = 75\ \mathrm{Mm}$, and $x = 0\ \mathrm{Mm}$ planes.}
    \label{fig:rhoT-early}
\end{figure*}

\begin{figure*}
    \centering
    \includegraphics[width=0.9\linewidth]{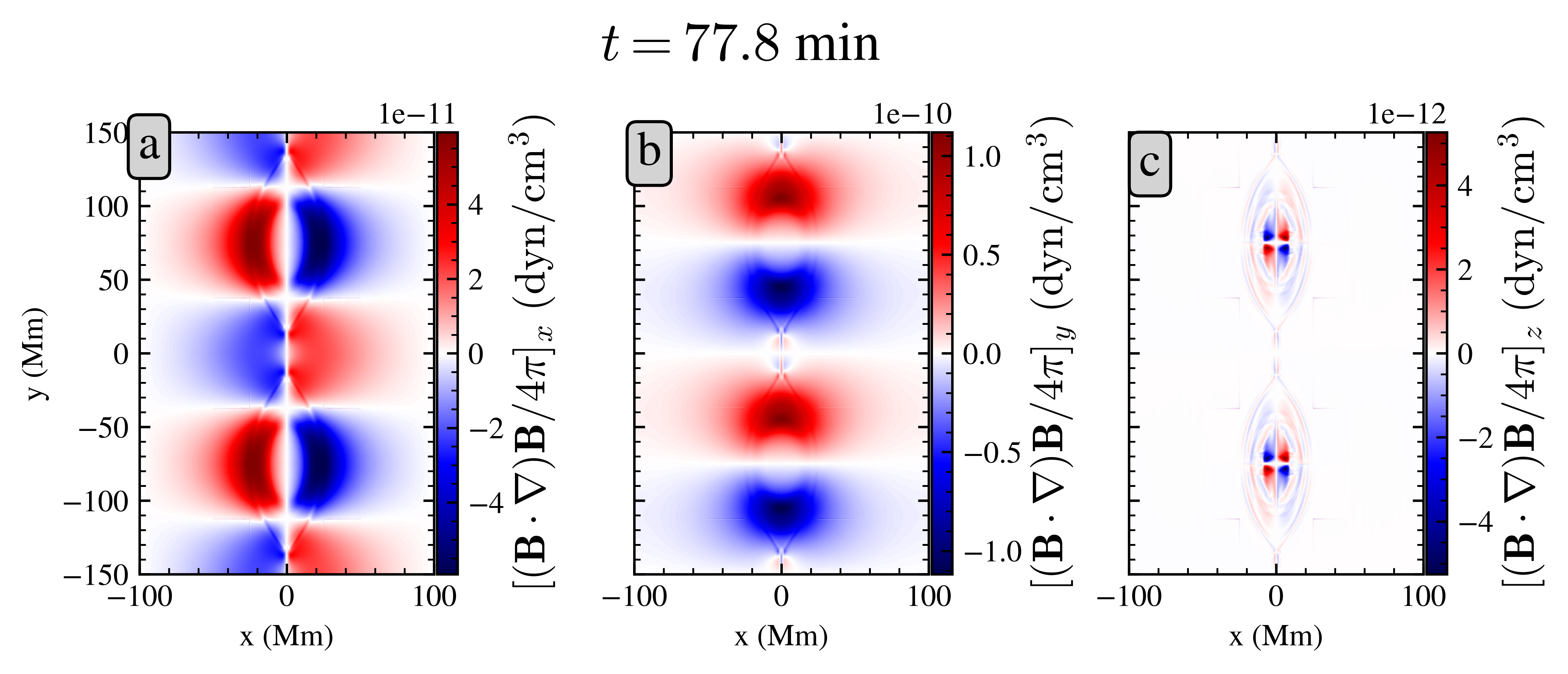}
    \caption{The components of the magnetic tension at $t = 77.8\ \mathrm{min}$, $z = 0\ \mathrm{Mm}$.}
    \label{fig:tension}
\end{figure*}

To further illustrate how density and temperature are related in this process, they are shown together in Fig. \ref{fig:rhoT-early}, at an early time ($t = 14.3\ \mathrm{min}$) in various planes, where we see a clear inverse correlation between density and temperature in the current sheet, with lower temperature coinciding with higher density and vice versa. When the thermal instability is triggered inside the current sheet near the flux rope axes (at around $t=80$ min), condensation sites draw in surrounding plasma. This is driven by the thermal pressure gradient and magnetic tension, shown at $t = 77.8\ \mathrm{min}$ in Fig. \ref{fig:tension}, whilst it is opposed by the magnetic pressure. This leads to the decrease of plasma density of the surrounding elliptical patch regions of the flux ropes. As our steady background heating is based on the initial density and temperature, there is a net heating in those patch regions which enhances the temperature. Indeed, this is reflected in Figs. \ref{fig:rhoT-end}a and \ref{fig:rhoT-end}d, where the hottest regions correspond to the lowest density regions. Between the hottest regions lies the cool plasmoid axis, with dense wings of a similar temperature extending outwards perpendicular to the current sheet.

\begin{figure*}
    \centering
    \includegraphics[width=0.9\linewidth]{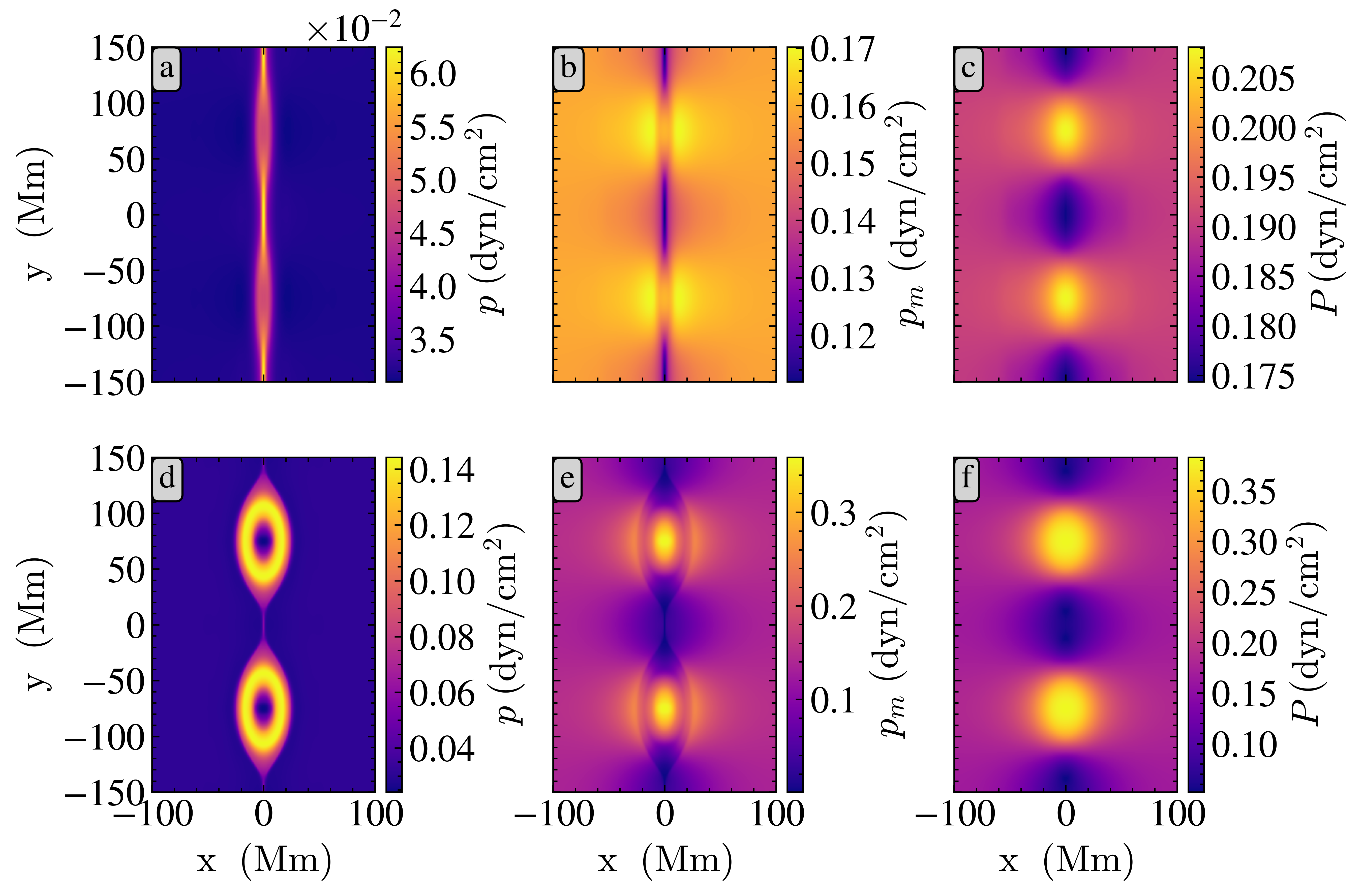}
    \caption{(a,d) Thermal pressure $p$, (b,e) magnetic pressure $p_m$, and (c,f) total pressure $P$ at $z = 0$, (a-c) $t = 14.3\ \mathrm{min}$ and (d-f) $t = 77.8\ \mathrm{min}$.}
    \label{fig:p-comparison}
\end{figure*}

As a final note on Figs. \ref{fig:rhoT-early} and \ref{fig:rhoT-end}, observe from the middle and right columns that there is no variation along the $z$-axis (this can also be seen in Fig. \ref{fig:isosurface-J2}). Though we established in Sec. \ref{sec:lin-dom} that the dominant wave vector is parallel to the $y$-axis, i.e. $\bfk_\text{max} \simeq (2\pi/15)\,\ey$ (dimensionless), the difference in growth rate with e.g. $\bfk = \bfk_\text{max} + (2\pi/30)\,\ez$ is only marginal (for comparison in Fig. \ref{fig:kvar}a, note that $\log_{10} (2\pi/30) \simeq -1.56$). Hence, even though the simulation was initiated with this dominant instability according to equation (\ref{eq:fourier}), and thus lacking a variation in $z$ because $k_3 = 0$, one might expect mixing to occur and a $z$-variation to develop. However, since a value of $k_3 = 2\pi/30$ has a wavelength in the $z$-direction equal to the box dimension, its development should definitely be noticeable as a periodic pattern from edge to edge, compatible with the periodic boundary conditions. Since no such thing is observed, we must again conclude that such mixing does not occur without deliberate breaking of the symmetry in the $z$-direction.

To show the role of the pressure in the density increase at the condensation sites, Fig. \ref{fig:p-comparison} shows the thermal pressure $p = n k_\mathrm{B} T$, the magnetic pressure $p_m = \bfb^2/8\pi$, and the total pressure $P = p+p_m$ at $z = 0$ for $t = 14.3\ \mathrm{min}$ and $t = 77.8\ \mathrm{min}$. Whereas the early stage features high thermal pressure at the plasmoid centre and strong magnetic pressure at the plasmoid edges, the condensation (and post-condensation) stage displays the opposite behaviour (high thermal pressure at the edges, strong magnetic pressure at the centre). Furthermore, in the early stage thermal pressure constitutes about a quarter of the total pressure whereas its contribution is significantly reduced compared to magnetic pressure later in the evolution. Therefore, even though the thermal pressure force points inwards, the overall pressure force is aimed outwards. Hence, it is the magnetic tension that is responsible for the mass migration to the plasmoid centre, where it causes the observed condensations.

\begin{figure*}
    \centering
    \includegraphics[width=0.9\linewidth]{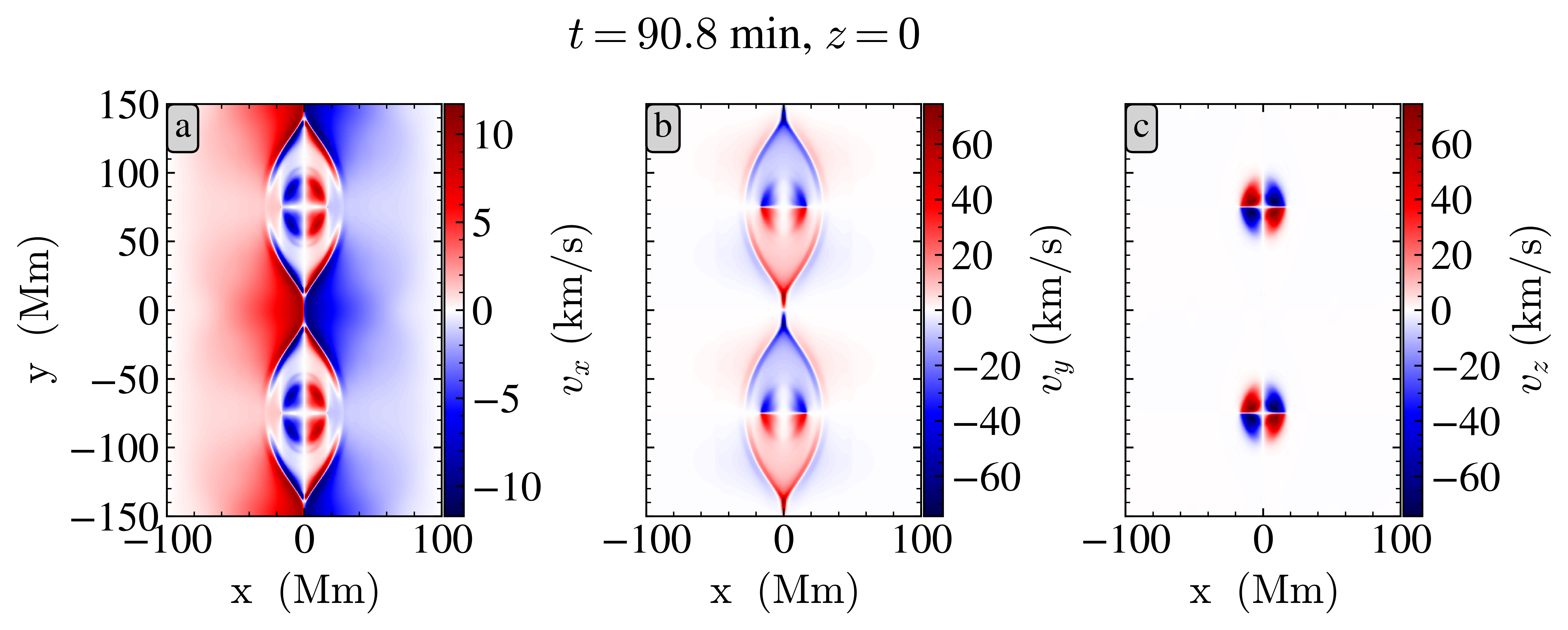}
    \caption{(a) $v_x$, (b) $v_y$, and (c) $v_z$ components at $t = 90.8\ \mathrm{min}$ in the $z = 0\ \mathrm{Mm}$ plane.}
    \label{fig:v-slices}
\end{figure*}

Returning to the velocity, of which it was already noted in the discussion of Fig. \ref{fig:t-evol}b that the $z$-component grows notably at the condensation time, Fig. \ref{fig:v-slices} now displays each component in the $z=0$ plane after condensation formation, at $t = 90.8\ \mathrm{min}$. Here, it becomes clear that the interior of each plasmoid decomposes into four quadrants, each with their own unique velocity direction. Calculating the angle between $\bfv$ and $\bfb$ reveals that close to the plasmoid axis and at the plasmoid edges, the flow is anti-parallel to $\bfb$ in the first and third quadrant, in the view of Fig. \ref{fig:v-slices}c, and parallel in the second and fourth quadrant. In a layer between this axial region and the plasmoid edge, the direction of flow deviates from this magnetic field alignment. The quadrupolar structure in the $v_z$-component is due to the magnetic tension since neither the thermal nor the magnetic pressure varies along $z$. As expected, this pattern is thus clearly present in the magnetic tension force at $t = 77.8\ \mathrm{min}$, where $|v_z|$ starts to increase, as shown in Fig. \ref{fig:temp-evol}b. As such, it is a signature of reconnection-driven flow, also seen in the 3D simulation of plasmoid instability without thermal effects \citep[see e.g.][]{Wang2015}. For the other two components, it is evident from comparing Figs. \ref{fig:p-comparison}d and \ref{fig:p-comparison}e to Fig. \ref{fig:p-comparison}f that the thermal and magnetic pressure counteract each other near the plasmoid edges. Furthermore, pressure (thermal and magnetic) and magnetic tension also work against each other and the balance between the two results in the observed $v_x$- and $v_y$-profiles. Note though that a less refined version of the velocity profile in Fig. \ref{fig:v-slices} (all components) is already present in the linear tearing solution, as shown in Fig. \ref{fig:initial}d-f.

\begin{figure}
    \centering
    \includegraphics[width=\linewidth]{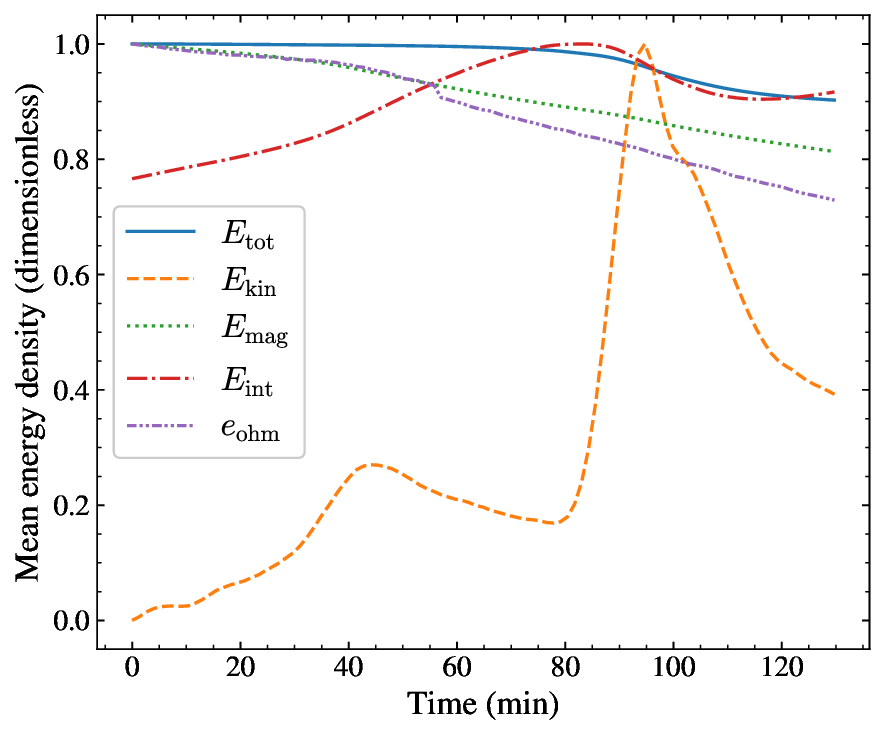}
    \caption{Evolution of the mean energy densities in the simulation, each normalised to their maximum value: $E_\mathrm{tot} = 2.07\times 10^{-1}\ \mathrm{erg}\,\mathrm{cm}^{-3}$, $E_\mathrm{kin} = 4.16\times 10^{-4}\ \mathrm{erg}\,\mathrm{cm}^{-3}$, $E_\mathrm{mag} = 1.59\times 10^{-1}\ \mathrm{erg}\,\mathrm{cm}^{-3}$, $E_\mathrm{int} = 6.23\times 10^{-2}\ \mathrm{erg}\,\mathrm{cm}^{-3}$, and $e_\mathrm{ohm} = 9.25\times 10^{-5}\ \mathrm{erg}\,\mathrm{cm}^{-3}\,\mathrm{s}^{-1}$.}
    \label{fig:E-evol}
\end{figure}

Finally, to justify the earlier statement on energy conversion at the reconnection sites, we look at the simulation's energy distribution. In order to do so, we first define the mean kinetic ($E_\mathrm{kin}$), magnetic ($E_\mathrm{mag}$), and internal ($E_\mathrm{int}$) energies as
\begin{align}
    E_\mathrm{kin} &= \frac{1}{V} \iiint_V \frac{\rho\bfv^2}{2}\dV, \\
    E_\mathrm{mag} &= \frac{1}{V} \iiint_V \frac{\bfb^2}{2}\dV, \\
    E_\mathrm{int} &= \frac{1}{V} \iiint_V \frac{p}{\gamma-1}\dV,
\end{align}
with $V$ the simulation box volume. The mean total energy is then given by the sum, $E_\mathrm{tot} = E_\mathrm{kin} + E_\mathrm{mag} + E_\mathrm{int}$. The total energy of the system is not affected by resistivity and thermal conduction, but a heating/cooling misbalance can lead to net energy gains or losses, depending on which effect dominates. Due to the inclusion of resistivity, magnetic energy is converted to internal energy at a rate of
\begin{equation}
    e_\mathrm{ohm} = \frac{1}{V} \iiint_V \eta\bfj^2\dV
\end{equation}
by Ohmic dissipation. The evolution of all of these quantities is shown in Fig. \ref{fig:E-evol}, where each curve is normalised to its maximum, found in the figure description. As the current sheet reconnects constantly throughout the simulation, the mean magnetic energy is decreasing monotonically, even past the thermal runaway process, contrary to the simulation in \citet{Sen2023}, where the magnetic energy was observed to increase again during the condensation formation. This discrepancy is likely due to the lack of symmetry breaking in the $z$-direction in the simulation here, resulting in a helical magnetic field structure here rather than the twisted structure in \citet{Sen2023}. The major jump in $v_z$ during the thermal runaway is also reflected in the kinetic energy here.

\begin{figure}
    \centering
    \includegraphics[width=\linewidth]{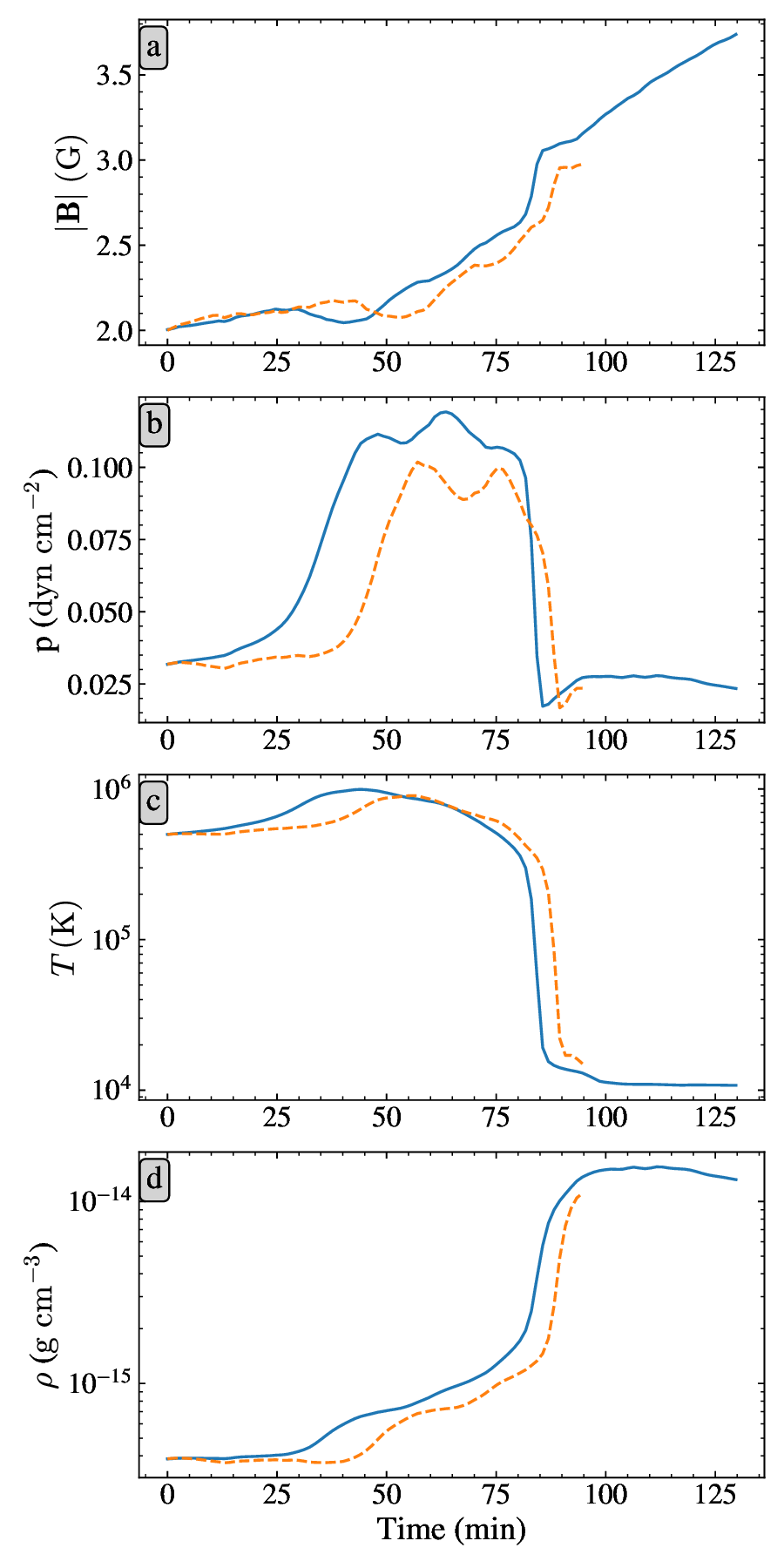}
    \caption{Evolution of the (a) magnetic field strength, (b) thermal pressure, (c) temperature, and (d) density, in the condensation region in simulations initialised with the tearing instability (solid blue) and thermal quasi-continuum modes (dashed orange).}
    \label{fig:time-series}
\end{figure}

Aside from the simulation initialised with the dominant (tearing) instability (Case 1), we also ran a simulation initialised with the superposition of the three thermal quasi-continuum modes shown in Figs. \ref{fig:ef}b and \ref{fig:ef}c (Case 2). However, aside from a half-period ($75\ \mathrm{Mm}$) shift in the $y$-direction, Case 2 showed remarkable similarity to Case 1, so we do not discuss it here in detail. Instead, Fig. \ref{fig:time-series} traces the magnetic field strength, thermal pressure, temperature, and density at a point in the condensation region in each simulation, namely $(10, 75, 0)\ \mathrm{Mm}$ and $(10, 0, 0)\ \mathrm{Mm}$, respectively. Case 1 results are shown with solid blue lines whereas Case 2 results are depicted with dashed orange lines, where we ran the simulation until we achieved the thermal runaway process. It is immediately clear that they exhibit very similar behaviour in the temporal evolution with only a small time delay of around 5 minutes for Case 2 compared to Case 1. This is presumably due to the difference in growth rate of the initial perturbation, before mode mixing kicks in.

\section{Conclusion}\label{sec:discuss}
To understand the magnetic, thermal, and dynamical behaviour of the solar corona, this work looked at the linear and non-linear stage of multimodal evolution, particularly in a coronal current sheet. This provides a theoretical basis for understanding the formation of plasmoids (magnetic flux ropes in 3D), which are observed in coronal current sheets \citep[][and references therein]{Chae:2017, Patel:2020}, as well as the formation of cool-condensations, which share similarities to the thermodynamic properties of the coronal rain observed in e.g. \citet{Qiao2024} forming in a current sheet that emerges between open and closed magnetic structures. Concretely, we focused on the stability properties and evolution of a force-free Harris current sheet in a fully ionized hydrogen plasma under solar coronal conditions. In a linear study with the \legolas{} code, it is revealed that this configuration's spectrum of eigenfrequencies features a (discrete) resistive tearing mode as well as an unstable thermal continuum. The resistivity has a destabilising effect on the thermal continuum, turning it into a quasi-continuum and smoothing out the continuum modes' discontinuities. For this specific setup, the resistive tearing mode was found to dominate, with the most unstable wave vector parallel to the direction of magnetic field inversion and a wavelength in the transition region between the constant- and nonconstant-$\psi$ classifications from the tearing literature.

To explore the non-linear evolution, the configuration was implemented as a 3D simulation in \amrvac{}, where a resistive tearing mode solution, close to the maximal growth rate and calculated by \legolas{}, was used as the initial perturbation. Initially, the tearing instability was observed to grow without any discernible mode mixing, i.e. without any influence of the unstable thermal modes and without any developing variation in the $z$-direction. Once the plasmoids had grown significantly, condensation started to form after $\sim 80\ \mathrm{min}$, first along the plasmoid axis and subsequently expanding towards the plasmoid edges perpendicularly to the current sheet. Notably, this is a lot sooner than in the simulation by \citet{Sen2023}, where condensation appeared after $\sim 140\ \mathrm{min}$. This may be due to their nonzero helium abundance, or initialisation with an artificial magnetic field perturbation, unlike our case, where the selected perturbations are the fastest growing modes in the linear stage. However, our second simulation, initialised with unstable thermal modes, did not reach the condensation stage much later than the simulation initialised with the tearing mode. This suggests that the instability growth rate of the current sheet in the non-linear phase, and achievement of the thermal runaway phase is solely governed during the non-linear evolution in a coupled tearing-thermal fashion, irrespective of the choice of tearing or thermal quasi-continuum modes at the linear (initial) stage. Furthermore, in a realistic configuration the initial perturbation will generally be a superposition of modes, with local differences. Variations in the timescale on which condensations form at specific sites may thus provide a clue to the properties of the local, initial perturbation.

Considering the linear growth rate is not very sensitive to the $z$-component of the wave vector, it is worth initialising this configuration with a tearing mode with a nonzero wave vector $z$-component to see how the structure along the plasmoid axis is affected. Naturally, another step forward then is to use a helium abundance appropriate for the solar corona, and compare that spectrum and evolution to this case and the simulation by \citet{Sen2023}. The two flux ropes in this simulation could also be forced to merge, e.g. by introducing velocity pinching at different locations along the current sheet \citep{Popescu:2023}, before or during the condensation process, which would definitely reveal an even more intricate condensation process. Furthermore, the current setup assumes a uniform coronal medium. How the instability growth rates change with the incorporation of a stratified medium with lower atmosphere (transition region and chromosphere) coupling also needs to be investigated, though this requires us to go beyond \legolas{}'s 1D capabilities. Nevertheless, the current work implies that initialising with the fastest growing linear mode skips the mode mixing phase in which the dominant instability traditionally emerges. This can significantly reduce the time-scale from the equilibrium phase to the non-linear stage and thermal runaway process. On the technical side, this reduces the computation time substantially.

\section*{Acknowledgements}
JDJ acknowledges funding by the UK's Science and Technology Facilities Council (STFC) Consolidated Grant ST/W001195/1. SS acknowledges support by the European Research Council through the Synergy Grant \#810218 (`The Whole Sun', ERC-2018-SyG).

%%%%%%%%%%%%%%%%%%%%%%%%%%%%%%%%%%%%%%%%%%%%%%%%%%
\section*{Data Availability}
A selection of \amrvac{} simulation snapshots that were visualised in this article, can be found in \citet{data}, alongside the \legolas{} data, and extracted time series data shown in Figs. \ref{fig:t-evol}, \ref{fig:E-evol}, and \ref{fig:time-series}. The data was obtained with \legolas{} v2.1.1 and \amrvac{} v3.1, which can be found at \url{https://legolas.science} and \url{https://amrvac.org}, respectively. Simulation analysis and visualisation were performed with \textsf{yt} \citep[\url{https://yt-project.org}]{Turk2011}.

% The inclusion of a Data Availability Statement is a requirement for articles published in MNRAS. Data Availability Statements provide a standardised format for readers to understand the availability of data underlying the research results described in the article. The statement may refer to original data generated in the course of the study or to third-party data analysed in the article. The statement should describe and provide means of access, where possible, by linking to the data or providing the required accession numbers for the relevant databases or DOIs.



\bibliographystyle{mnras}
\bibliography{bibliography} % if your bibtex file is called example.bib

%\begin{thebibliography}{99}
%\end{thebibliography}

%%%%%%%%%%%%%%%%%%%%%%%%%%%%%%%%%%%%%%%%%%%%%%%%%%

%%%%%%%%%%%%%%%%% APPENDICES %%%%%%%%%%%%%%%%%%%%%

% \appendix

% \section{Some extra material}

% If you want to present additional material which would interrupt the flow of the main paper,
% it can be placed in an Appendix which appears after the list of references.

%%%%%%%%%%%%%%%%%%%%%%%%%%%%%%%%%%%%%%%%%%%%%%%%%%

% Don't change these lines
\bsp	% typesetting comment
\label{lastpage}
\end{document}